\documentclass[journal]{IEEEtran}
\usepackage{amsthm}  %provides the environment for proof
\usepackage{amsfonts}
\usepackage{color}
\usepackage[normalem]{ulem}
\usepackage{graphicx}
\usepackage[tight,footnotesize]{subfigure}
\usepackage{psfrag}
\usepackage{subfigure}
\usepackage{epsfig}
\usepackage{amssymb}
\usepackage{tabu}
\usepackage{cite} %文献显示格式[1]-[3]

\usepackage{algorithmic,algorithm}%算法

\makeatother
\usepackage{multirow}
\usepackage[cmex10]{amsmath}
\usepackage{bm}%zhou
\newcounter{rcounter}
\allowdisplaybreaks[4]

\newtheorem{remark}{Remark}
\newtheorem{theorem}{Theorem}

\newtheorem{lemma}{Lemma}

\newtheorem{corollary}{Corollary}

\newtheorem{proposition}{Proposition}

\def\ScaleIfNeeded{%
\ifdim\Gin@nat@width>\linewidth \linewidth \else \Gin@nat@width
\fi } \makeatother

\begin{document}

\title{Non-orthogonal Multiple Access in Large-Scale Heterogeneous Networks}
%When Massive MIMO Meets HetNets in Large-Scale Networks: is NOMA Practical?}
%
%\author{
%\IEEEauthorblockN{  Yuanwei~Liu\IEEEauthorrefmark{1}, Zhijin~Qin\IEEEauthorrefmark{1}, Maged Elkashlan\IEEEauthorrefmark{1}, Yue~Gao\IEEEauthorrefmark{1}, and Arumugam Nallanathan\IEEEauthorrefmark{2} } \IEEEauthorblockA{
%%\{z.qin, yuanwei.liu, yue.gao, maged.elkashlan\}@qmul.ac.uk\\
%%\ arumugam.nallanathan@kcl.ac.uk\\
% } }

 \author{
  Yuanwei~Liu,~\IEEEmembership{Member,~IEEE,}
        Zhijin~Qin,~\IEEEmembership{Member,~IEEE,}
        Maged~Elkashlan,~\IEEEmembership{Member,~IEEE,}
        Arumugam~Nallanathan,~\IEEEmembership{Fellow,~IEEE,}
        and Julie~A.~McCann,~\IEEEmembership{Member,~IEEE},

\thanks{Part of this work has been presented in IEEE Global Communication Conference (GLOBECOM),  Dec. Washington D.C, USA, 2016 \cite{yuanwei2016Hetnets}.}
\thanks{Y. Liu and A. Nallanathan are with the Department of Informatics, King's College London, London WC2R 2LS, U.K. (email: \{yuanwei.liu, arumugam.nallanathan\}@kcl.ac.uk).}
\thanks{Z. Qin and J. McCann are with the Department of Computing, Imperial College London, London SW7 2AZ,
U.K. (email: \{z.qin,  j.mccann\}@imperial.ac.uk).}
\thanks{M. Elkashlan is with the School of Electronic Engineering and Computer Science, Queen Mary University of London, London E1 4NS,
U.K. (email: maged.elkashlan@qmul.ac.uk).}

}

\maketitle
\begin{abstract}
In this paper, the potential benefits of applying non-orthogonal multiple access (NOMA) technique in $K$-tier hybrid heterogeneous networks (HetNets) is explored. A  promising new transmission framework is proposed, in which NOMA is adopted in small cells and massive multiple-input multiple-output (MIMO) is employed in macro cells. For maximizing the biased average received power for mobile users, a NOMA and massive MIMO based user association scheme is developed.  To evaluate the performance of the proposed framework, we first derive the analytical expressions for the coverage probability of NOMA enhanced small cells. We then examine the spectrum efficiency of the whole network, by deriving exact analytical expressions for NOMA enhanced small cells and a tractable lower bound for massive MIMO enabled macro cells. Lastly, we investigate the energy efficiency of the hybrid HetNets.   Our results demonstrate that: 1) The coverage probability of NOMA enhanced small cells is affected to a large extent by the targeted transmit rates and power sharing coefficients of two NOMA users;  2) Massive MIMO enabled macro cells are capable of significantly enhancing the spectrum efficiency by increasing the number of antennas; 3) The energy efficiency of the whole network can be greatly improved by densely deploying NOMA enhanced small cell base stations (BSs); and 4) The proposed NOMA enhanced HetNets transmission scheme has superior performance compared to the orthogonal multiple access~(OMA) based HetNets.
\end{abstract}

%Stochastic geometry approaches are invoked to model the locations of BSs and users in the considered large-scale networks.

\section{Introduction}
The last decade has witnessed the escalating data explosion on the Internet \cite{CiscoVNI}, which is brought by the emerging demanding applications such as high-definition videos, online games and virtual reality. Also, the rapid development of internet of things (IoT) requires for facilitating billions of devices to communicate with each other \cite{IOT2015}. Such requirements pose new challenges for designing the fifth-generation (5G) networks. Driven by these challenges, non-orthogonal multiple access (NOMA), a promising technology for 5G networks, has  attracted  much attention for its potential ability to enhance spectrum efficiency~\cite{Dai2015NOMA} and improving user access~\cite{shirvanimoghaddam2016massive,Zhijin2017modulation}. The key idea of NOMA\footnote{In this treatise, we use ``NOMA" to refer to ``power-domain NOMA"
for simplicity.} is to utilize a superposition coding (SC) technology at the transmitter and successive interference cancelation (SIC) technology at the receiver \cite{Zhiguo2015Mag}, and hence multiple access can be realized in power domain via different power levels for users in the same resource block.  Some initial research investigations have been made in this field \cite{Saito:2013,ding2014performance,Timotheou:2015,Jinho:2015}.  The system-level performance of the downlink NOMA with two users has been demonstrated in \cite{Saito:2013}. In \cite{ding2014performance}, the performance of a general NOMA transmission has been evaluated in which one base station (BS) is able to communicate with several spatial randomly deployed users.  As a further advance, the fairness issue of NOMA has been addressed in \cite{Timotheou:2015},  by examining appropriate power allocation policies among the NOMA users. For multi-antenna NOMA systems, a two-stage multicast beamforming downlink transmission scheme has been proposed in~\cite{Jinho:2015}, where the total transmitter power was optimized using closed-form expressions.

Heterogeneous networks (HetNets) and massive multiple-input multiple-output (MIMO), as two ``big three" technologies~\cite{andrews2014will}, are seen as the fundamental structure for the 5G networks. The core idea of HetNets is to establish closer BS-user links by densely overlaying small cells. By doing so, promising benefits such as lower power consumption, higher throughput and enhanced spectrum spatial reuse can be experienced~\cite{Andrews2008Femto}. The massive MIMO regime enables tens of hundreds/thousands antennas at a BS, and hence it is capable of offering an unprecedented level of freedom to serve multiple mobile users \cite{Xie2016access}. Aiming to fully take advantage of both massive MIMO and HetNets, in \cite{Adhikary2015Hetnets}, interference coordination issues found in massive MIMO enabled HetNets was addressed  by utilizing the spatial blanking of macro cells.  In \cite{ye2015user}, the authors investigated a joint user association and interference management optimization problem in massive MIMO HetNets.

%As a further advance, considering the unbiased  the coverage provability of the massive MIMO enabled heterogeneous networks was examined in

\subsection{Motivation and Related Works}
Sparked by the aforementioned potential benefits, we therefore explore the potential performance~enhancement brought by NOMA for the hybrid HetNets. Stochastic geometry is an effective mathematical tool for capturing the topological randomness of networks. As such, it is capable of providing tractable analytical results in terms of average network behaviors~\cite{chiu2013stochastic}. Some research contributions with utilizing stochastic geometry approaches have been studied in the context of Hetnets and NOMA~\cite{jo2012heterogeneous,Dhillon2014MIMOHetnets,Wen2016Hetnets,Anqi2015Hetnets,yuanwei_JSAC_2015,Zhiguo2016General,Yuanwei2017TWC}. For HetNets scenarios, based on applying a flexible bias-allowed user association approach, the performance of multi-tier downlink HetNets has been examined in \cite{jo2012heterogeneous}, where all BSs and users were assumed to be equipped with a single antenna.  As a further advance, the coverage provability of the multi-antenna enabled HetNets has been investigated in \cite{Dhillon2014MIMOHetnets}, using a simple selection bias based cell selection policy. By utilizing massive MIMO enabled HetNets and a stochastic geometry model, the spectrum efficiency of uplinks and downlinks  were evaluated in \cite{Wen2016Hetnets} and \cite{Anqi2015Hetnets}, respectively.

Regarding the literature of stochastic geometry based NOMA scenarios, an incentive user cooperation NOMA protocol was proposed in \cite{yuanwei_JSAC_2015} to tackle spectrum and energy issues, by regarding near users as energy harvesting relays for improving the reliability of far users. By utilizing signal alignment technology, a new MIMO-NOMA design framework has been proposed in a stochastic geometry based model \cite{Zhiguo2016General}.  Driven by the security issues, two effective approaches---protection zone and artificial noise has been utilized to enhance the physical layer security for NOMA in large-scale networks in \cite{Yuanwei2017TWC}. Very recently, the potential co-existence of two technologies, NOMA and millimeter wave (mmWave) has been examined in \cite{Zhiguo2016mmwave}, in which the random beamforming technology is adopted.

Despite the ongoing research contributions having played a vital role for fostering HetNets and NOMA technologies, to the best of our knowledge, the impact of NOMA enhanced hybrid HetNets design has not been researched.  Also, there is lack of complete systematic performance evaluation metrics, i.e., coverage probability and energy efficiency. Different from the conventional HetNets design \cite{jo2012heterogeneous,Wen2016Hetnets}, NOMA enhanced HetNets design poses three additional challenges: i) NOMA technology brings additional co-channel interference from the superposed signal of the connected BS; ii) NOMA technology requires careful channel ordering design to carry out SIC operations at the receiver; and iii) the user association policy requires  consideration of power sharing effects of NOMA. Aiming at tackling the aforementioned issues, developing a systematic mathematically tractable framework for intelligently investigating the effect of various types of interference on network performance is desired.
\subsection{Contributions and Organization}
We propose a new hybrid HetNets framework with NOMA enhanced small cells and massive MIMO aided macro cells. We believe that the proposed structure design can contribute to the design of a more promising 5G system due to the following key advantages:
\begin{itemize}
  \item High spectrum efficiency: With higher BS densities, the NOMA enhanced BSs are capable of accessing the served users closer, which increase the transmit signal-to-interference-plus-noise ratio (SINR) by intelligently tracking multi-category interference, such as inter/intra-tier interference and intra-BS interference.
  \item Low complexity: By applying NOMA in single-antenna based small cells, the complex cluster based precoding/detection design for MIMO-NOMA systems \cite{ding2015mimo,Yuanwei2016fairness} can be avoided.
  \item Fairness/throughput tradeoff: NOMA is capable of addressing fairness issues by allocating more power to weak users~\cite{Zhiguo2015Mag}, which is of great significance for HetNets when investigating efficient resource allocation in sophisticated large-scale multi-tier networks.
\end{itemize}

%NOMA is regarded as a promising ``add-on" technology for the existing OMA based networks due to the gradually mature of SC and SIC technologies, and will not bring much implementation complexity.

Different from  most existing stochastic geometry based single cell research contributions in terms of NOMA \cite{ding2014performance,yuanwei_JSAC_2015,Zhiguo2016General,Yuanwei2017TWC,Zhiguo2016mmwave}, we consider multi-cell multi-tier scenarios in this treatise, which is more challenging. In this framework, we consider a downlink $K$-tier HetNets, where macro BSs are equipped with large antenna arrays with linear zero-forcing beamforming (ZFBF) capability to serve multiple single-antenna users simultaneously, and small cells BSs equipped with single antenna each to serve two single-antenna users simultaneously with NOMA transmission. Based on the proposed design, the primary theoretical contributions are summarized as follows:
 %we provide a tractable analytical framework to characterize the performance of the considered networks.
\begin{enumerate}
  \item We develop a flexible biased association policy to address the impact of NOMA and massive MIMO on the maximum biased received power. Utilizing this policy, we first derive the exact analytical expressions for the coverage probability of a typical user associating with the NOMA enhanced small cells for the most general case. Additionally, we derive closed-form expressions in terms of coverage probability for the interference-limited case that each tier has the same path loss.
  \item We derive the exact analytical expressions of the NOMA enhanced small cells in terms of spectrum efficiency. Regarding the massive MIMO enabled macro cells, we provide a tractable analytical lower bound for the most general case and closed-form expressions for the case that each tier has the same path loss. Our analytical results illustrate that the spectrum efficiency can be greatly enhanced by increasing the scale of large antenna arrays.
  \item We finally derive the energy efficiency of the whole network by applying a popular power consumption model \cite{power2014massive}. Our results reveal that NOMA enhanced small cells achieve higher energy efficiency than macro cells. It is also shown that increasing antenna numbers at the macro cell BSs has the opposite effect on energy efficiency.
  \item We show  that the NOMA enhanced small cell design has superior performance over conventional orthogonal multiple access (OMA) based small cells in terms of coverage probability, spectrum efficiency and energy efficiency, which demonstrates the benefits of the proposed framework.
\end{enumerate}

The rest of the paper is organized as follows. In Section II, the network model for NOMA enhanced hybrid HetNets is introduced. In Section III, new analytical expressions  for the coverage probability of the NOMA enhanced small cells are derived. Then spectrum efficiency and energy efficiency are investigated in Section IV and Section V, respectively. Numerical results are presented in Section VI, which is followed by the conclusions in Section VII.

\section{Network Model}

\subsection{Network Description}

\begin{figure}[t!]
    \begin{center}
        \includegraphics[width=3.5in]{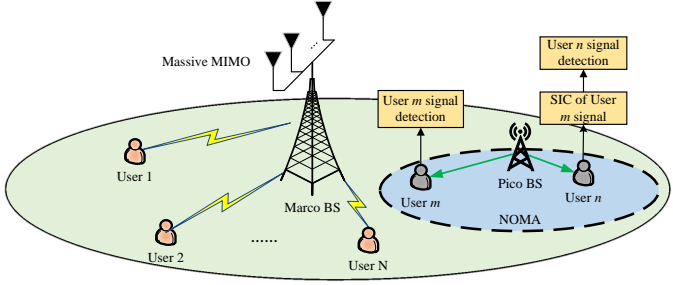}
        \caption{Illustration of NOMA and massive MIMO based hybrid HetNets.}
        \label{System model}
    \end{center}
\end{figure}

Focusing on downlink transmission scenarios, we consider a $K$-tier HetNets model, where the first tier represents the macro cells and the other tiers represent the small cells, such as pico cells and femto cells. The positions of macro BSs and all the $k$-th tier $\left( {k \in \left\{ {2, \cdots ,K} \right\}} \right)$ BSs are modeled as homogeneous poisson point processes (HPPPs) ${\Phi _1}$ and ${\Phi _k}$ and with density ${\lambda _1}$ and ${\lambda _k}$, respectively. As it is common to overlay a high-power macro  cell with successively denser and lower power small cells, we apply massive MIMO technologies to macro cells and NOMA to small cells in this work. As shown in Fig.~\ref{System model}, in macro cells, BSs are equipped with $M$ antennas, each macro BS transmits signals to $N$ users over the same resource block (e.g., time/frequency/code). We assume that $M \gg N > 1$ and linear ZFBF technique is applied at each macro BS assigning equal power to $N$ data streams~\cite{huh2012network}. Perfect downlink channel state information (CSI) are assumed at the BSs. In small cells, each BS is equipped with single antenna. Such structure consideration is to avoid sophisticated MIMO-NOMA precoding/detection in small cells. All users are considered to be equipped with single antenna each.  We adopt user pairing in each tier of small cells to implement NOMA to lower the system complexity \cite{yuanwei_JSAC_2015}. It is worth pointing out that in long term evolution advanced (LTE-A), NOMA also implements a form of two-user case \cite{3GPP}.
\subsection{NOMA and Massive MIMO Based User Association}
%We assume that each small cell BS has  already served one user in a random distance within a disc, the radius of  disc for the $k$-th tier BSs is set as $r_k$. Then we squeeze one more user to each BS, and each BS is supposed to be able to communicate with its connected two users by applying NOMA protocol. As a result,
In this work,  a user is allowed to access the BS of any tier, which provides the best coverage. We consider flexible user association based on the maximum average received power of each tier.  %\textcolor[rgb]{0.00,0.00,1.00}{We assume that each small cell BS has  already connected with one user. Then we squeeze one more user to each small cell BS, and each BS is supposed to be able to communicate with its connected two users by applying NOMA protocol.}

\subsubsection{Average received power in NOMA enhanced small cells} Different from the convectional user association in OMA, NOMA exploits the power sparsity for multiple access by allocating different powers to different users. Due to the random spatial topology of the stochastic geometry model, the space information of users are not pre-determined.  The user association policy for the NOMA enhanced small cells assumes that a near user is chosen as the typical one first. As such, at the $i$-th tier small cell, the averaged  power received at users connecting to the $i$-th tier BS $j$ (where $ j \in {\Phi _i}$) is given by:
\begin{align}\label{received power small cell}
{P_{r,i}} = {a_{n,i}}{P_i}L\left( {{d_{j,i}}} \right){B_i},
\end{align}
where ${P_i}$ is the transmit power of a $i$-th tier BS, $a_{n,i}$ is the power sharing coefficient for the near user, $L\left( {{d_{j,i}}} \right) = \eta d_{j,i}^{ - {\alpha _i}}$ is large-scale path loss, ${d_{j,i}}$ is the distance between the user and a $i$-th tier BS, $\alpha_i$ is the path loss exponent of the $i$-th tier small cell, $\eta $ is the frequency dependent factor, and $B_i$ is the identical bias factor which are useful for offloading data traffic in HetNets.

\subsubsection{Average received power in massive MIMO aided macro cells} In macro cells, as the macro BS is equipped with multiple antennas,  macro cell users experience large array gains. By adopting the ZFBF transmission scheme, the array gain obtained at macro users is ${G_M} = M - N + 1$~\cite{huh2012network,hosseini2014large}. As a result, the average  power received at users connecting to macro BS $\ell$ (where $ \ell  \in {\Phi _M}$) is given by
%\vspace{-0.2cm}
\begin{align}\label{received power Macro}
{P_{r,1}} = {G_M}{P_1}L\left( {{d_{\ell ,1}}} \right)/N,
\end{align}
where ${P_1 }$ is the transmit power of a macro BS, $L\left( {{d_{\ell ,1}}} \right) = \eta d_{\ell ,1}^{ - {\alpha _1}}$ is the large-scale path loss, ${d_{\ell ,1}}$ is the distance between the user and a macro BS.

\subsection{Channel Model}

\subsubsection{NOMA enhanced small cell transmission} In small cells, without loss of generality, we consider that each small cell BS is associated with one user in the previous round of user association process. Applying the NOMA protocol, we aim to squeeze a typical user into a same small cell to improve the spectral efficiency. For simplicity, we assume that the distances between the associated users and the connected small cell BSs are the same, which can be arbitrary values and are denoted as $r_k$, future work will relax this assumption. The distance between a typical user and the connected small cell BS is random.  Due to the fact that the path loss is more stable and dominant compared to the instantaneous small-scale fading \cite{steele1999mobile}, we assume that the SIC operation always happens at the near user. We denote that  ${d_{o,{k_m}}}$ and ${d_{o,{k_n}}}$ are the distances from the $k$-th tier small cell BS to user $m$ and user $n$, respectively. Since it is not pre-determined that a typical user is a near user $n$ or a far user $m$, we have the following near user case and far user case.

\textbf{Near user case}: When a typical user has smaller distance to the BS than the connected user ($x \le r_k$, here $x$ denotes the distance between the typical user and the BS), then we have ${d_{o,{k_m}}}=r_k$. Here we use $m^*$ to represent the user which has been already connected to the BS in the last round of user association process, we use $n$ to represent the typical user in near user case.  User $n$  will first decode the information of the connected user $m^*$ to the same BS with the following SINR
\begin{align}\label{SINR small cell kn m near case}
{\gamma _{{k_{n \to m*}}}} = \frac{{{a_{m,k}}{P_k}{g_{o,k}}L\left( {{d_{o,{k_n}}}} \right)}}{{{a_{n,k}}{P_k}{g_{o,k}}L\left( {{d_{o,k}}} \right) + {I_{M,k}} + {I_{S,k}} + {\sigma ^2}}},
\end{align}
where $a_{m,k}$ and $a_{n,k}$ are the power sharing coefficients for two users in the $k$-th layer, ${\sigma ^2}$ is the additive white Gaussian noise (AWGN) power, $L\left( {{d_{o,{k_n}}}} \right) = \eta d_{o,{k_n}}^{ - {\alpha _i}}$ is the large-scale path loss, ${I_{M,k}} = \sum\nolimits_{\ell  \in {\Phi _1}} {\frac{{{P_1}}}{N}{g_{\ell ,1}}L\left( {{d_{\ell ,1}}} \right)}$ is the interference from macro cells, ${I_{S,k}} = \sum\nolimits_{i = 2}^K {{\sum _{j \in {\Phi _i}\backslash {B_{o,k}}}}{P_i}{g_{j,i}}L\left( {{d_{j,i}}} \right)}$ is the interference from small cells,  ${{g_{o,{{k}}}}}$ and ${{d_{o,{k_n}}}}$ refer the small-scale fading coefficients and distance between a typical user and the BS in the $k$-th tier, ${{g_{\ell ,1}}}$ and ${{d_{\ell ,1}}}$ refer the small-scale fading coefficients and distance between a typical user and  BS $\ell$ in the macro cell, respectively, ${{g_{j ,i}}}$ and ${{d_{j ,i}}}$ refers to the small-scale fading coefficients and distance between a typical user and its connected  BS $j$ except the serving BS ${B_{o,k}}$ in the $i$-th tier small cell, respectively. Here, ${{g_{o,k}}}$ and ${{g_{j ,i}}}$ follow exponential distributions with unit mean. ${{g_{\ell ,1}}}$  following Gamma distribution with parameters $\left( {N ,1} \right)$.

If the information of user $m^*$ can be decoded successfully, user $n$ then decodes its own message.  As such, the SINR at a typical user $n$, which connects with the $k$-th tier small cell, can be expressed as
\begin{align}\label{SINR small cell kn near case}
{\gamma _{{k_n}}} = \frac{{{a_{n,k}}{P_k}{g_{o,k}}L\left( {{d_{o,{k_n}}}} \right)}}{{{I_{M,k}} + {I_{S,k}} + {\sigma ^2}}}.
\end{align}

For the connected far user $m^*$ served by the same BS, the signal can be decoded by treating the message of user $n$ as interference. Therefore, the SINR that for the connected user $m^*$ to the same BS in the $k$-th tier small cell can be expressed as
\begin{align}\label{SINR small cell rm near case}
{\gamma _{{k_{m^*}}}} = \frac{{{a_{m,k}}{P_k}{g_{o,k}}L\left( {{r_k}} \right)}}{{{I_{k,n}} + {I_{M,k}} + {I_{S,k}} + {\sigma ^2}}},
\end{align}
where ${I_{k,n}} = {a_{n,k}}{P_k}{g_{o,k}}L\left( {{r_{k}}} \right)$, and $L\left( {{r_k}} \right) = \eta {r_k}^{ - {\alpha _k}}$.

\textbf{Far user case:}  When a typical user has a larger distance to the BS than the connected user($x > r_k$), we have ${d_{o,{k_n}}}=r_k$. Here we use $n^*$ to represent the user which has been already connected to the BS in the last round of user association process, we use $m$ to represent the typical user in far user case. As such, for the connected near user $n^*$, it will first decode the information of user $m$ with the following SINR
\begin{align}\label{SINR small cell rn m far case}
{\gamma _{{k_{n* \to m}}}} = \frac{{{a_{m,k}}{P_k}{g_{o,k}}L\left( {{r_k}} \right)}}{{{a_{n,k}}{P_k}{g_{o,k}}L\left( {{r_k}} \right) + {I_{M,k}} + {I_{S,k}} + {\sigma ^2}}}.
\end{align}

Once user $m$ is decoded successfully, the interference from a typical user $m$ can be canceled, by applying SIC technique. Therefore, the SINR at the connected user $n^*$ to the same BS  in the $k$-th tier small cell is given by
\begin{align}\label{SINR small cell rn far case}
{\gamma _{{k_{n^*}}}} = \frac{{{a_{n,k}}{P_k}{g_{o,k}}L\left( {{r_k}} \right)}}{{{I_{M,k}} + {I_{S,k}} + {\sigma ^2}}}.
\end{align}

For user $m$ that connects to the $k$-th tier small cell, the SINR can be expressed as
\begin{align}\label{SINR small cell km far case}
{\gamma _{{k_m}}} = \frac{{{a_{m,k}}{P_k}{g_{o,k}}L\left( {{d_{o,{k_m}}}} \right)}}{{{I_{k,n^*}} + {I_{M,k}} + {I_{S,k}} + {\sigma ^2}}},
\end{align}
where ${I_{k,n^*}} = {a_{n,k}}{P_k}{g_{o,k}}L\left( {{d_{o,{k_m}}}} \right)$, $L\left( {{d_{o,{k_m}}}} \right) = \eta d_{o,{k_m}}^{ - {\alpha _k}}$, ${{d_{o,{k_n}}}}$ is the distance between a typical user $m$ and the connected BS in the $k$-th tier.

\subsubsection{Massive MIMO aided macro cell transmission} Without loss of generality, we assume that a typical user is located at the origin of an infinite two-dimension plane. Based on \eqref{received power small cell} and \eqref{received power Macro}, the SINR at a typical user that connects with a macro BS at a random distance ${d_{o,1}}$ can be expressed as

\begin{align}\label{SINR Macro}
{\gamma _{r,1}} = \frac{{\frac{{{P_1}}}{N}{h_{o,1}}L\left( {{d_{o,1}}} \right)}}{{{I_{M,1}} + {I_{S,1}} + {\sigma ^2}}},
\end{align}
where ${I_{M,1}} = \sum\nolimits_{\ell  \in {\Phi _1}\backslash {B_{o,1}}} {\frac{{{P_1}}}{N}{h_{\ell ,1}}L\left( {{d_{\ell ,1}}} \right)}$ is the interference from the macro cells, ${I_{S,1}} = \sum\nolimits_{i = 2}^K {\sum\nolimits_{j \in {\Phi _i}} {{P_i}{h_{j,i}}L\left( {{d_{j,i}}} \right)} } $ is the interference from the small cells;  ${{h_{o,1}}}$ is the small-scale fading coefficient between a typical user and the connected macro BS, ${{h_{\ell ,1}}}$ and ${{d_{\ell ,1}}}$ refer to the small-scale fading coefficients and distance between a typical user and the connected macro BS $\ell$ except for the serving BS ${B_{o,1}}$ in the macro cell, respectively, ${{h_{j ,i}}}$ and ${{d_{j ,i}}}$ refer to the small-scale fading coefficients and distance between a typical user and BS $j$ in the $i$-th tier small cell, respectively. Here, ${{h_{o,1}}}$  follows Gamma distribution with parameters $\left( {M - N + 1,1} \right)$, ${{h_{\ell ,1}}}$  follows Gamma distribution with parameters $\left( {N ,1} \right)$, and ${{h_{j ,i}}}$ follows exponential distribution with unit mean.

\section{Coverage Probability of Non-orthogonal Multiple Access Based Small Cells}
In this section, we focus our attention on analyzing the coverage probability of a typical user associated to the NOMA enhanced small cells, which is different from the conventional OMA based small cells due to the channel ordering of two users.  % The analysis of the coverage probability of a typical user associated with macro cells is the same as the conventional OMA small cells, which has been investigated in \cite{Anqi2015GC} and hence is skipped in this treatise.
\subsection{User Association Probability and Distance Distributions}
As described in Section II-B, the user association of the proposed framework is based on maximizing the biased average received power at the users. As such, based on \eqref{received power small cell}  and \eqref{received power Macro}, the user association of macro cells and small cells are given by the following. For simplicity, we denote ${{\tilde B}_{ik}} = \frac{{{B_i}}}{{{B_k}}}, {{\tilde \alpha }_{ik}} = \frac{{{\alpha _i}}}{{{\alpha _k}}},{{\tilde \alpha }_{1k}} = \frac{{{\alpha _1}}}{{{\alpha _k}}},{{\tilde \alpha }_{i1}} = \frac{{{\alpha _i}}}{{{\alpha _1}}}$, ${{\tilde P}_{1k}} = \frac{{{P_1}}}{{{P_k}}},{{\tilde P}_{i1}} = \frac{{{P_i}}}{{{P_1}}}$, and ${{\tilde P}_{ik}} = \frac{{{P_i}}}{{{P_k}}}$ in the following parts of this work.

\begin{lemma}\label{User association}
The user association probability that a typical user connects to the NOMA enhanced small cell BSs in the $k$-th tier and to the macro BSs can be calculated as:
\begin{align}\label{UA probability k}
{A_k} =& 2\pi {\lambda _k}\int_0^\infty  {r\exp \left[ { - \pi \sum\limits_{i = 2}^K {{\lambda _i}{{\left( {{{\tilde P}_{ik}}{{\tilde B}_{ik}}} \right)}^{{\delta _i}}}{r^{\frac{2}{{{{\tilde \alpha }_{ik}}}}}}} } \right.}  \nonumber\\
&\left. { - \pi {\lambda _1}{{\left( {\frac{{{{\tilde P}_{1k}}{G_M}}}{{N{a_{n,k}}{B_k}}}} \right)}^{{\delta _1}}}{r^{\frac{2}{{{{\tilde \alpha }_{1k}}}}}}} \right]dr.,
\end{align}
and
\begin{align}\label{UA probability macro}
{A_1} =& 2\pi {\lambda _1}\int_0^\infty  {r\exp \left[ { - \pi {{\sum\limits_{i = 2}^K {{\lambda _i}\left( {\frac{{{a_{n,i}}{{\tilde P}_{i1}}{B_i}N}}{{{G_M}}}} \right)} }^{{\delta _i}}}{r^{\frac{2}{{{{\tilde \alpha }_{i1}}}}}}} \right.} \nonumber\\
&\left. { - \pi {\lambda _1}{r^2}} \right]dr.,
\end{align}
respectively, where ${\delta _1} = \frac{2}{{{\alpha _1}}}$ and ${\delta _i} = \frac{2}{{{\alpha _i}}}$.
\begin{proof}
Using a similar method to Lemma 1 of~\cite{jo2012heterogeneous}, \eqref{UA probability k} and  \eqref{UA probability macro} can be easily obtained.
\end{proof}
\end{lemma}

%Based on \eqref{UA probability macro}, the number of users per macro BS is give by
%\begin{align}\label{User number macro}
%{{\mathcal{N}}_1} = \frac{{{\mathcal{A}_1}{\lambda _u}}}{{{\lambda _1}}},
%\end{align}

%Based on \eqref{UA probability k}, the number of users per small cell BS in the $k$-th tier is give by
%\begin{align}\label{User number k}
%{{\mathcal{N}}_k} = \frac{{{\mathcal{A}_k}{\lambda _u}}}{{{\lambda _k}}}.
%\end{align}
\begin{corollary}\label{User association}
For the special case that each tier has the same path loss exponent, i.e., ${\alpha _1} = {\alpha _k} = \alpha $, the user association probability of the NOMA enhanced small cells in the $k$-th tier and the macro cells can be expressed in closed form as
\begin{align}\label{UA probability k special}
{{\tilde A}_k} = \frac{{{\lambda _k}}}{{\sum\limits_{i = 2}^K {{\lambda _i}{{\left( {{{\tilde P}_{ik}}{{\tilde B}_{ik}}} \right)}^\delta }}  + {\lambda _1}{{\left( {\frac{{{{\tilde P}_{1k}}{G_M}}}{{N{a_{n,k}}{B_k}}}} \right)}^\delta }}},
\end{align}
and
\begin{align}\label{UA probability macro special}
{{\tilde A}_1} = \frac{{{\lambda _1}}}{{\sum\limits_{i = 2}^K {{\lambda _i}{{\left( {\frac{{{a_{n,i}}{{\tilde P}_{i1}}{B_i}N}}{{{G_M}}}} \right)}^\delta }}  + {\lambda _1}}},
\end{align}
respectively, where ${\delta} = \frac{2}{{{\alpha}}}$.
\end{corollary}

\begin{remark}\label{User association remark}
The derived results in \eqref{UA probability k special}  and \eqref{UA probability macro special} demonstrate that by increasing the number of antennas at the macro cell BSs, the user association probability of the macro cells increases and the user association probability of the small cells decreases. This is due to the large array gains brought by the macro cells to the  users served. It is also worth noting that increasing the power sharing coefficient, $a_n$, results in a higher association probability of small cells. As $a_n\rightarrow1$, the user association becomes the same as in the conventional OMA based approach.
\end{remark}

We consider the probability density function (PDF) of the distance between a typical user and its connected  small cell BS in the $k$-th tier. Based on \eqref{UA probability k}, we obtain
\begin{align}\label{PDF distance_k}
{f_{{d_{o,k}}}}\left( x \right) =& \frac{{2\pi {\lambda _k}x}}{{{A_k}}}\exp \left[ { - \pi \sum\limits_{i = 2}^K {{\lambda _i}{{\left( {{{\tilde P}_{ik}}{{\tilde B}_{ik}}} \right)}^{{\delta _i}}}{x^{\frac{2}{{{{\tilde \alpha }_{ik}}}}}}} } \right.\nonumber\\
&{ - \pi {\lambda _1}{{\left( {\frac{{{{\tilde P}_{1k}}{G_M}}}{{N{a_{n,k}}{B_k}}}} \right)}^{{\delta _1}}}{x^{\frac{2}{{{{\tilde \alpha }_{1k}}}}}}}.
\end{align}

We then calculate the PDF of the distance between a typical user and its connected macro BS. Based on \eqref{UA probability macro}, we obtain
\begin{align}\label{PDF distance_macro}
{f_{{d_{o,1}}}}\left( x \right) =& \frac{{2\pi {\lambda _1}x}}{{{A_1}}}\exp \left[ { - \pi {{\sum\limits_{i = 2}^K {{\lambda _i}\left( {\frac{{{a_{n,i}}{{\tilde P}_{i1}}{B_i}N}}{{{G_M}}}} \right)} }^{{\delta _i}}}{x^{\frac{2}{{{{\tilde \alpha }_{i1}}}}}}} \right.\nonumber\\
&\left. { - \pi {\lambda _1}{x^2}} \right].
\end{align}

\subsection{The Laplace Transform of Interference}
The next step is to derive the Laplace transform of a typical user. We denote that ${I_{k}} = {I_{S,k}} +{I_{M,k}}$ is the total interference to the typical user in the $k$-th tier. The laplace transform of $I_{k}$ is ${\mathcal{L}_{{I_{k}}}}\left( s \right) = {\mathcal{L}_{{I_{S,k}}}}\left( s \right){\mathcal{L}_{{I_{M,k}}}}\left( s \right)$. We first calculate the Laplace transform of interference from the small cell BS to a typical user $ {{\cal L}_{I_{S,k}}}\left( s \right) $ in the following Lemma.
\begin{lemma}\label{Typical Laplace conventional NOMA small}
The Laplace transform of interference from the small cell BSs to a typical user can be expressed as
\begin{align}\label{Laplace_near_typical_2}
{{\cal L}_{I_{S,k}}}\left( s \right) =& \exp \left\{ { - s\sum\limits_{i = 2}^K {\frac{{{\lambda _i}2\pi {P_i}\eta {{\left( {{\omega _{i,k}}\left( {{x_0}} \right)} \right)}^{2 - {\alpha _i}}}}}{{{\alpha _i}\left( {1 - {\delta _i}} \right)}} \times } } \right.\nonumber\\
&\left. {{}_2{F_1}\left( {1,1 - {\delta _i};2 - {\delta _i}; - s{P_i}\eta {{\left( {{\omega _{i,k}}\left( {{x_0}} \right)} \right)}^{ - {\alpha _i}}}} \right)} \right\},
\end{align}
where ${}_2{F_1}\left( {\cdot,\cdot;\cdot;\cdot} \right)$ is the is the Gauss hypergeometric function \cite[ Eq. (9.142)]{gradshteyn}, and ${\omega _{i,k}}\left( {{x_0}} \right) = {\left( {{{\tilde B}_{ik}}{{\tilde P}_{ik}}} \right)^{\frac{{{\delta _i}}}{2}}}x_0^{\frac{1}{{{{\tilde \alpha }_{ik}}}}}$ is the nearest distance allowed between the typical user and its connected small cell BS in the $k$-th tier. %Here, we denote ${{\tilde B}_{ik}} = \frac{{{B_i}}}{{{B_k}}},{{\tilde P}_{ik}} = \frac{{{P_i}}}{{{P_k}}}$, and ${{\tilde \alpha }_{ik}} = \frac{{{\alpha _i}}}{{{\alpha _k}}}$.

\begin{proof}
See Appendix A.
\end{proof}
\end{lemma}
Then we  calculate the laplace transform of interference from the macro cell to a typical user ${\mathcal{L}_{{I_{M,k}}}}\left( s \right)$ in the following Lemma.
\begin{lemma}\label{Typical Laplace conventional NOMA macro}
The Laplace transform of interference from the macro cell BSs to a typical user can be expressed as
\begin{align}\label{Lapalace M k final}
&{\mathcal{L}_{{I_{M,k}}}}\left( s \right) = \exp \left[ { - {\lambda _1}\pi {\delta _1}\sum\limits_{p = 1}^N {
N\choose
p} {{\left( {s\frac{{{P_1}}}{N}\eta } \right)}^p}{{\left( { - s\frac{{{P_1}}}{N}\eta } \right)}^{{\delta _1} - p}}} \right.\nonumber\\
&\left. { \times B\left( { - s\frac{{{P_1}}}{N}\eta {{\left[ {{\omega _{1,k}}\left( x_0 \right)} \right]}^{ - {\alpha _1}}};p - {\delta _1},1 - N} \right)} \right],
\end{align}
where ${B}\left( {\cdot;\cdot,\cdot} \right)$ is the is the incomplete Beta function \cite[ Eq. (8.319)]{gradshteyn},  and ${\omega _{1,k}}\left( {{x_0}} \right) = {\left( {\frac{{{{\tilde P}_{1k}}{G_M}}}{{{a_{n,k}}{B_k}N}}} \right)^{\frac{{{\delta _1}}}{2}}}{x^{\frac{1}{{{{\tilde \alpha }_{1k}}}}}}$ is the nearest distance allowed between a typical user and its connected BS in the macro cell. %Here we denote ${{\tilde P}_{1k}} = \frac{{{P_1}}}{{{P_k}}}$ and ${{\tilde \alpha }_{1k}} = \frac{{{\alpha _1}}}{{{\alpha _k}}}$
\begin{proof}
See Appendix B.
\end{proof}
\end{lemma}

\subsection{Coverage Probability}
The coverage probability is defined as that a typical user can successfully transmit signals with a targeted data rate $R_t$. According to the distances, two cases are considered in the following.

\textbf{Near user case:} For the near user case, ${{x_0}}< r_k$, successful decoding will happen when the following  two conditions hold:
\begin{enumerate}
  \item The typical user can decode the message of the connected user served by the same BS.
  \item After the SIC process, the typical user can decode its own message.
\end{enumerate}

As such, the coverage probability of the typical user on the condition of the distance ${x_0}$ in the $k$-th tier is:
\begin{align}\label{successful_probability_near_typical_1}
{\left. {{P_{cov,k}}\left( {{\tau _c},{\tau _t},{x_0}} \right)} \right|_{{x_0} \le {r_k}}} = \Pr \left\{ {{\gamma _{{k_{n \to m*}}}} > {\tau _c},{\gamma _{{k_n}}} > {\tau _t}} \right\},
\end{align}
where ${\tau _t} = {2^{{R_t}}} - 1$ and ${\tau _c} = {2^{{R_c}}} - 1$. Here ${R_c}$ is the targeted data rate of the connected user served by the same BS.

Based on \eqref{successful_probability_near_typical_1}, for the near user case, we can obtain the expressions for the conditional coverage probability of a typical user  in the following Lemma.
\begin{lemma}\label{Sucessful typical near}
If ${a_{m,k}} - {\tau _c}{a_{n,k}} \ge 0$ holds, the conditional coverage probability of a typical user for the near user case is expressed in closed-form as
\begin{align}\label{successful_probability_near_typical_3}
&{\left. {{P_{cov,k}}\left( {{\tau _c},{\tau _t},{x_0}} \right)} \right|_{{x_0} \le {r_k}}} = \exp \left\{ { - \frac{{{\varepsilon ^*}\left( {{\tau _c},{\tau _t}} \right)x_0^{{\alpha _k}}{\sigma ^2}}}{{{P_k}\eta }}} \right.\nonumber\\
& - {\lambda _1}{\delta _1}\pi {\left( {{{\tilde P}_{1k}}{\varepsilon ^*}\left( {{\tau _c},{\tau _t}} \right)/N} \right)^{{\delta _1}}}x_0^{\frac{2}{{{{\tilde \alpha }_{1k}}}}}Q_{1,t}^n\left( {{\tau _c},{\tau _t}} \right)\nonumber\\
&\left. { - \sum\limits_{i = 2}^K {\frac{{{\lambda _i}{\delta _i}\pi {{\left( {{{\tilde B}_{ik}}} \right)}^{\frac{2}{{{\alpha _i}}} - 1}}{{\left( {{{\tilde P}_{ik}}} \right)}^{\frac{2}{{{\alpha _i}}}}}x_0^{\frac{2}{{{{\tilde \alpha }_{ik}}}}}}}{{1 - {\delta _i}}}Q_{i,t}^n\left( {{\tau _c},{\tau _t}} \right)} } \right\}.
\end{align}
Otherwise, ${\left. {{P_{cov,k}}\left( {{\tau _c},{\tau _t},{x_0}} \right)} \right|_{{x_0} \le {r_k}}}=0$. Here, $\varepsilon _t^n = \frac{{{\tau _t}}}{{{a_{n,k}}}}$, $\varepsilon _c^f = \frac{{{\tau _c}}}{{{a_{m,k}} - {\tau _c}{a_{n,k}}}}$, ${\varepsilon ^*}\left( {{\tau _c},{\tau _t}} \right) = \max \left\{ {\varepsilon _c^f,\varepsilon _t^n} \right\}$, $Q_{i,t}^n\left( {{\tau _c},{\tau _t}} \right) = {\varepsilon ^*}\left( {{\tau _c},{\tau _t}} \right){}_2{F_1}\left( {1,1 - {\delta _i};2 - {\delta _i}; - \frac{{{\varepsilon ^*}\left( {{\tau _c},{\tau _t}} \right)}}{{{{\tilde B}_{ik}}}}} \right)$, and  $Q_{1,t}^n\left( {{\tau _c},{\tau _t}} \right) = \sum\limits_{p = 1}^N {
N\choose
p}{{\left( { - 1} \right)}^{{\delta _1} - p}} \times$\\
$B\left( { - \frac{{{\varepsilon ^*}\left( {{\tau _c},{\tau _t}} \right){a_{n,k}}{B_k}}}{{{G_M}}};p - {\delta _1},1 - N} \right)$.
\begin{proof}
Substituting \eqref{SINR small cell kn m near case} and \eqref{SINR small cell kn near case} into \eqref{successful_probability_near_typical_1}, we obtain
\begin{align}\label{successful_probability_near_typical_2}
&{\left. {{P_{cov,k}}\left( {{\tau _c},{\tau _t},{x_0}} \right)} \right|_{{x_0} \le {r_k}}} = \Pr \left\{ {\frac{{{g_{o,k_n}}{P_k}\eta }}{{x_0^{{\alpha _i}}\left( {{I_k} + {\sigma ^2}} \right)}} > {\varepsilon ^*}\left( {{\tau _c},{\tau _t}} \right)} \right\}\nonumber\\
&= {e^{ - \frac{{{\varepsilon ^*}\left( {{\tau _c},{\tau _t}} \right)x_0^{{\alpha _k}}{\sigma ^2}}}{{{P_k}\eta }}}}{\mathbb{E}_{{I_k}}}\left\{ {{e^{ - \frac{{{\varepsilon ^*}\left( {{\tau _c},{\tau _t}} \right)x_0^{{\alpha _k}}y}}{{{P_k}\eta }}}}} \right\}\nonumber\\
&= {e^{ - \frac{{{\varepsilon ^*}\left( {{\tau _c},{\tau _t}} \right)x_0^{{\alpha _k}}{\sigma ^2}}}{{{P_k}\eta }}}}{{\cal L}_{{I_k}}}\left( {\frac{{{\varepsilon ^*}\left( {{\tau _c},{\tau _t}} \right)}}{{{P_k}\eta }}x_0^{{\alpha _k}}} \right).
\end{align}
Then by plugging \eqref{Laplace_near_typical_2} and \eqref{Lapalace M k final} into \eqref{successful_probability_near_typical_2}, we obtain the conditional coverage probability for the near user case in \eqref{successful_probability_near_typical_3}. The proof is complete.
\end{proof}
\end{lemma}

%\begin{lemma}\label{Sucessful connect near}
%Conditioned on ${{x_0} \le {r_k}}$, the successful probability of a connect user for the near user case is expressed as
%\begin{align}\label{successful_probability connect_near_typical_3}
%&{\left. {{P_{cov,k}}\left( {{\tau _c},{\tau _t},{x_0}} \right)} \right|_{{x_0} \le {r_k}}} = \exp \left\{ { - \frac{{{\varepsilon ^*}\left( {{\tau _c},{\tau _t}} \right)x_0^{{\alpha _k}}{\sigma ^2}}}{{{P_k}\eta }}} \right.\nonumber\\
%&\left. { - \sum\limits_{i = 2}^K {\frac{{{\lambda _i}{\delta _i}\pi {{\left( {{{\tilde B}_{ik}}} \right)}^{\frac{2}{{{\alpha _i}}} - 1}}{{\left( {{{\tilde P}_{ik}}} \right)}^{\frac{2}{{{\alpha _i}}}}}x_0^{\frac{2}{{{{\tilde \alpha }_{ik}}}}}}}{{1 - {\delta _i}}}Q_{i,t}^n\left( {{\tau _c},{\tau _t}} \right)} } \right\},
%\end{align}
%\begin{proof}
%\end{proof}
%\end{lemma}

\textbf{Far user case:} For the far user case, ${{x_0}}> r_k$, successful decoding will happen if the typical user can decode its own message by treating the connected user served by the same BS as noise. The conditional coverage probability of a typical user for the far user case is calculated in the following Lemma.

\begin{lemma}\label{Sucessful typical far}
If ${a_{m,k}} - {\tau _t}{a_{n,k}} \ge 0$ holds, the coverage probability of a typical user for the far user case is expressed in closed-form as
\begin{align}\label{successful_probability_far_typical_1}
&{\left. {{P_{cov,k}}\left( {{\tau _t},{x_0}} \right)} \right|_{{x_0} > {r_k}}} = \exp \left\{ { - \frac{{\varepsilon _t^fx_0^{{\alpha _k}}{\sigma ^2}}}{{{P_k}\eta }}} \right.\nonumber\\
 & - {\lambda _1}{\delta _1}\pi {\left( {{{\tilde P}_{1k}}\varepsilon _t^f/N} \right)^{{\delta _1}}}x_0^{\frac{2}{{{{\tilde \alpha }_{1k}}}}}Q_{1,t}^f\left( {{\tau _t}} \right)\nonumber\\
&\left. { - \sum\limits_{i = 2}^K {\frac{{{\lambda _i}{\delta _i}\pi {{\left( {{{\tilde B}_{ik}}} \right)}^{\frac{2}{{{\alpha _i}}} - 1}}{{\left( {{{\tilde P}_{ik}}} \right)}^{\frac{2}{{{\alpha _i}}}}}x_0^{\frac{2}{{{{\tilde \alpha }_{ik}}}}}}}{{1 - {\delta _i}}}Q_{i,t}^f\left( {\tau _t} \right)} } \right\}.
\end{align}
Otherwise, ${\left. {{P_{cov,k}}\left( {{\tau _t},{x_0}} \right)} \right|_{{x_0} > {r_k}}}=0$. Here $\varepsilon _t^f = \frac{{{\tau _t}}}{{{a_{m,k}} - {\tau _t}{a_{n,k}}}}$, and $Q_{1,t}^f\left( {{\tau _t}} \right) = \sum\limits_{p = 1}^N {
N\choose
p}{{\left( { - 1} \right)}^{{\delta _1} - p}} B\left( { - \frac{{\varepsilon _t^f{a_{n,k}}{B_k}}}{{{G_M}}};p - {\delta _1},1 - N} \right)$ $Q_{i,t}^f\left( {\tau _t} \right) = \varepsilon _t^f{}_2{F_1}\left( {1,1 - {\delta _i};2 - {\delta _i}; - \frac{{\varepsilon _t^f}}{{{{\tilde B}_{ik}}}}} \right)$.
\begin{proof}
Based on \eqref{SINR small cell km far case}, we have
\begin{align}\label{successful_probability_far_typical}
{\left. {{P_{cov,k}}\left( {{\tau _t},{x_0}} \right)} \right|_{{x_0} > {r_k}}} = \Pr \left\{ {{g_{o,{k_m}}} > \frac{{\varepsilon _t^fx_0^{{\alpha _i}}\left( {{I_k} + {\sigma ^2}} \right)}}{{{P_k}\eta }}} \right\}.
\end{align}
Following the similar procedure to obtain \eqref{successful_probability_near_typical_3}, with interchanging ${{\varepsilon ^*}\left( {{\tau _c},{\tau _t}} \right)}$ with ${\varepsilon _t^f}$, we obtain the desired results in \eqref{successful_probability_far_typical_1}. The proof is complete.
\end{proof}
\end{lemma}

Based on \textbf{Lemma \ref{Sucessful typical near}} and \textbf{Lemma \ref{Sucessful typical far}},  we can calculate the coverage probability of a typical user in the following Theorem.

\begin{theorem}\label{Sucessful typical total}
The coverage probability of a typical user associated to the $k$-th tier small cells is expressed as
\begin{align}\label{successful_probability_typical_final}
&{P_{cov,k}}\left( {{\tau _c},{\tau _t}} \right) = \int_0^{{r_k}} {{{\left. {{P_{cov,k}}\left( {{\tau _c},{\tau _t},{x_0}} \right)} \right|}_{{x_0} \le {r_k}}}} f_{{d_{o,k}}}\left( {{x_0}} \right)d{x_0}\nonumber\\
& + \int_{{r_k}}^\infty  {{{\left. {{P_{cov,k}}\left( {{\tau _t},{x_0}} \right)} \right|}_{{x_0} > {r_k}}}} f_{{d_{o,k}}}\left( {{x_0}} \right)d{x_0},
\end{align}
where ${{{\left. {{P_{cov,k}}\left( {{\tau _c},{\tau _t},{x_0}} \right)} \right|}_{{x_0} \le {r_k}}}} $ is given in \eqref{successful_probability_near_typical_3}, ${{{\left. {{P_{cov,k}}\left( {{\tau _t},{x_0}} \right)} \right|}_{{x_0} > {r_k}}}}$ is given in \eqref{successful_probability_far_typical_1}, and $f_{{d_{o,k}}}\left( {{x_0}} \right)$  is given in \eqref{PDF distance_k}.
\begin{proof}
Based on \eqref{successful_probability_near_typical_3} and \eqref{successful_probability_far_typical_1}, considering the distant distributions of a typical user associated to the $k$-th user small cells, we can easily obtain the desired results in \eqref{successful_probability_typical_final}. The proof is complete.
\end{proof}
\end{theorem}

Although \eqref{successful_probability_typical_final} has provided the exact analytical expression for the coverage probability of a typical user, it is difficult to directly obtain insights from this expression. Driven by this, we provide one special case that considers each tier with the same path loss exponents. As such, we have ${{{\tilde \alpha }_{1k}}}={{{\tilde \alpha }_{ik}}} = 1$.  In addition, we consider the interference limited case, where the thermal noise can be neglected\footnote{This is a common assumption in stochastic geometry based large-scale networks \cite{jo2012heterogeneous,liu2015secure}.}. Then based on \eqref{successful_probability_typical_final}, we can obtain the closed-form coverage probability of a typical user in the following Corollary.

%For simplicity, we consider the power allocation factor in terms of NOMA in each tier are the same, i.e., ${a_{n,k}} = {a_n}, {a_{m,k}} = {a_m}$.

\begin{corollary}\label{Sucessful typical special total}
With ${\alpha _1} = {\alpha _k} = \alpha $ and ${\sigma ^2} = 0$, the coverage probability of a typical user can be expressed in closed-form as follows:
\begin{align}\label{successful_probability_typical_special final}
{{\tilde P}_{cov,k}}\left( {{\tau _c},{\tau _t}} \right) = &\frac{{{b_k}\left( {1 - {e^{ - \pi \left( {{b_k} + c_1^n\left( {{\tau _c},{\tau _t}} \right) + c_2^n\left( {{\tau _c},{\tau _t}} \right)} \right)r_k^2}}} \right)}}{{{b_k} + c_1^n\left( {{\tau _c},{\tau _t}} \right) + c_2^n\left( {{\tau _c},{\tau _t}} \right)}}\nonumber\\
& + \frac{{{b_k}{e^{ - \pi \left( {{b_k} + c_1^f\left( {{\tau _t}} \right) + c_2^f\left( {{\tau _t}} \right)} \right)r_k^2}}}}{{{b_k} + c_1^f\left( {{\tau _t}} \right) + c_2^f\left( {{\tau _t}} \right)}},
\end{align}
where ${b_k} = \sum\limits_{i = 2}^K {{\lambda _i}{{\left( {{{\tilde P}_{ik}}{{\tilde B}_{ik}}} \right)}^\delta }}  + {\lambda _1}{\left( {\frac{{{{\tilde P}_{1k}}{G_M}}}{{N{a_{n,k}}{B_k}}}} \right)^\delta }$, $c_1^n\left( {{\tau _c},{\tau _t}} \right) = {\lambda _1}{\delta _1}{\left( {\frac{{{{\tilde P}_{1k}}{\varepsilon ^*}\left( {{\tau _c},{\tau _t}} \right)}}{N}} \right)^\delta }\tilde Q_{1,t}^n\left( {{\tau _c},{\tau _t}} \right)$, $c_2^n\left( {{\tau _c},{\tau _t}} \right) = \sum\limits_{i = 2}^K {\frac{{{\lambda _i}{\delta _i}{{\left( {{{\tilde B}_{ik}}} \right)}^{\frac{2}{\alpha } - 1}}{{\left( {{{\tilde P}_{ik}}} \right)}^{\frac{2}{\alpha }}}}}{{1 - {\delta _i}}}\tilde Q_{i,t}^n\left( {{\tau _c},{\tau _t}} \right)}$, $c_1^f\left( {{\tau _t}} \right) = {\lambda _1}{\delta _1}{\left( {\frac{{{{\tilde P}_{1k}}\varepsilon _t^f}}{N}} \right)^{{\delta _1}}}\tilde Q_{1,t}^f\left( {{\tau _t}} \right)$, and $c_2^f\left( {{\tau _t}} \right) = \sum\limits_{i = 2}^K {\frac{{{\lambda _i}{\delta _i}{{\left( {{{\tilde B}_{ik}}} \right)}^{\frac{2}{\alpha } - 1}}{{\left( {{{\tilde P}_{ik}}} \right)}^{\frac{2}{\alpha }}}}}{{1 - \delta }}\tilde Q_{i,t}^f\left( {{\tau _t}} \right)}$. Here, $\tilde Q_{1,t}^n\left( {{\tau _c},{\tau _t}} \right),\tilde Q_{i,t}^n\left( {{\tau _c},{\tau _t}} \right),\tilde Q_{1,t}^f\left( {{\tau _t}} \right)$, and  $\tilde Q_{i,t}^f\left( {{\tau _t}} \right)$ are based on interchanging the same path loss exponents, i.e. ${\alpha _1} = {\alpha _k} = \alpha $, for each tier from  $ Q_{1,t}^n\left( {{\tau _c},{\tau _t}} \right), Q_{i,t}^n\left( {{\tau _c},{\tau _t}} \right), Q_{1,t}^f\left( {{\tau _t}} \right)$, and  $ Q_{i,t}^f\left( {{\tau _t}} \right)$.
\begin{proof}
If ${\alpha _1} = {\alpha _k} = \alpha $ hols, \eqref{UA probability k} can be rewritten as
\begin{align}\label{Ak_special}
{{\tilde A}_k} = \frac{{{\lambda _1}}}{{{b_k} }},
\end{align}
Then we have
\begin{align}\label{f small_special}
{{\tilde f}_{{d_{o,k}}}}\left( x \right) = 2\pi {b_k}x\exp \left( { - \pi {b_k}{x^2}} \right).
\end{align}
Then by plugging \eqref{f small_special} into \eqref{successful_probability_typical_final} and after some mathematical manipulations, we can obtain the desired results in \eqref{successful_probability_typical_special final}.
\end{proof}
\end{corollary}

\begin{remark}\label{coverage small remark}
The derived results in \eqref{successful_probability_typical_special final} demonstrate that the coverage probability of a typical user is determined by both the target rate of itself and the target rate of the connected user served by the same BS. Additionally, inappropriate power allocation such as, ${a_{m,k}} - {\tau _t}{a_{n,k}} < 0$, will lead to the coverage probability always being zero.
\end{remark}

\section{Spectrum Efficiency}
To evaluate the spectrum efficiency of the proposed NOMA enhanced hybrid HetNets framework, we calculate the spectrum efficiency of each tier in this section.
\subsection{Ergodic Rate of NOMA enhanced Small Cells}
Rather than calculating the coverage probability of the case with fixed targeted rate, the achievable ergodic rate for NOMA enhanced small cells is opportunistically determined by the channel conditions of users. It is easy to verify that if the far user can decode the message of itself, the near user can definitely decode the message of far user since it has better channel conditions~\cite{ding2014performance}. Recall that the distance order between the connected BS and the two users are not predetermined, as such, we calculate the achievable ergodic rate of small cells both for the near user case and far user case in the following Lemmas.
\begin{lemma}\label{theorem:small near}
The achievable ergodic rate of the $k$-th tier small cell for the near user case can be expressed as follows:

\begin{align}\label{troughput small cell near}
\tau _k^n = \frac{{2\pi {\lambda _k}}}{{{A_k}\ln 2}}\left[ {\int_0^{\frac{{{a_{m,k}}}}{{{a_{n,k}}}}} {\frac{{{{\bar F}_{{\gamma _{{k_{m*}}}}}}\left( z \right)}}{{1 + z}}} dz + \int_0^\infty  {\frac{{{{\bar F}_{{\gamma _{{k_n}}}}}\left( z \right)}}{{1 + z}}} dz} \right],
\end{align}
where ${{{\bar F}_{{\gamma _{{k_{m*}}}}}}\left( z \right)}$ and ${{{\bar F}_{{\gamma _{{k_n}}}}}\left( z \right)}$ are given by

\begin{align}\label{CDF rm}
{{{\bar F}_{{\gamma _{{k_{m*}}}}}}\left( z \right)} =& \int_0^{{r_k}} {x\exp \left[ { - \frac{{{\sigma ^2}z{r_k}^{{\alpha _k}}}}{{\left( {{a_{m,k}} - {a_{n,k}}z} \right){P_k}\eta }}} \right.} \nonumber\\
&\left. { - \Theta \left( {\frac{{z{r_k}^{{\alpha _k}}}}{{\left( {{a_{m,k}} - {a_{n,k}}z} \right){P_k}\eta }}} \right) + \Lambda \left( x \right)} \right]dx,
\end{align}
and
\begin{align}\label{CDF kn}
&{{{\bar F}_{{\gamma _{{k_n}}}}}\left( z \right)} = \int_0^{{r_k}} {x\exp \left[ {\Lambda \left( x \right) - \frac{{{\sigma ^2}z{x^{{\alpha _k}}}}}{{{a_{n,k}}{P_k}\eta }} - \Theta \left( {\frac{{z{x^{{\alpha _k}}}}}{{{a_{n,k}}{P_k}\eta }}} \right)} \right]} dx.
\end{align}
Here $\Lambda \left( x \right) =  - \pi \sum\limits_{i = 2}^K {{\lambda _i}{{\left( {{{\tilde P}_{ik}}{{\tilde B}_{ik}}} \right)}^{{\delta _i}}}{x^{\frac{2}{{{{\tilde \alpha }_{ik}}}}}}}  - \pi {\lambda _1}{\left( {\frac{{{{\tilde P}_{1k}}{G_M}}}{{N{a_{n,k}}{B_k}}}} \right)^{{\delta _1}}}{x^{\frac{2}{{{{\tilde \alpha }_{1k}}}}}}$ and $\Theta \left( {s} \right)$ is given by
\begin{align}\label{laplace inter}
&\Theta \left( {s} \right) = {\lambda _1}\pi {\delta _1}\sum\limits_{p = 1}^N {
 N \choose
 p } {\left( {s\frac{{{P_1}}}{N}\eta } \right)^p}{\left( { - s\frac{{{P_1}}}{N}\eta } \right)^{{\delta _1} - p}}\nonumber\\
&\times B\left( { - s\frac{{{P_1}}}{N}\eta {{\left[ {{\omega _{1,k}}\left( x \right)} \right]}^{ - {\alpha _1}}};p - {\delta _1},1 - N} \right)\nonumber\\
&+ s\sum\limits_{i = 2}^K {\frac{{{\lambda _i}2\pi {P_i}\eta {{\left( {{\omega _{i,k}}\left( x \right)} \right)}^{2 - {\alpha _i}}}}}{{{\alpha _i}\left( {1 - {\delta _i}} \right)}}} \nonumber\\
&\times \left. {{}_2{F_1}\left( {1,1 - {\delta _i};2 - {\delta _i}; - s{P_i}\eta {{\left( {{\omega _{i,k}}\left( x \right)} \right)}^{ - {\alpha _i}}}} \right)} \right].
\end{align}
%where ${\omega _{1,k}}\left( x \right) = {\left( {\frac{{{{\tilde P}_{1k}}{G_M}}}{{{a_{n,k}}{B_k}N}}} \right)^{\frac{{{\delta _1}}}{2}}}{x^{\frac{1}{{{{\tilde \alpha }_{1k}}}}}}$ and ${\omega _{i,k}}\left( x \right) = {\left( {{{\tilde P}_{ik}}{{\tilde B}_{ik}}} \right)^{\frac{{{\delta _i}}}{2}}}{x^{\frac{1}{{{{\tilde \alpha }_{1k}}}}}}$ are the nearest distance allowed between the typical user associated to the $k$-th tier small cell and the macro cell BS, and between the typical user and the $i$-th tier small cell BS, respectively. ${B}\left( {\cdot;\cdot,\cdot} \right)$ and ${}_2{F_1}\left( {\cdot,\cdot;\cdot;\cdot} \right)$ are the the incomplete Beta function \cite[ Eq. (8.319)]{gradshteyn} and  Gauss hypergeometric function \cite[ Eq. (9.142)]{gradshteyn}, respectively.
\begin{proof}
See Appendix~C.
\end{proof}
\end{lemma}

\begin{lemma}\label{theorem:small far}
The achievable ergodic rate of the $k$-th tier small cell for the far user case can be expressed as follows:
\begin{align}\label{troughput small cell far}
\tau _k^f = \frac{{2\pi {\lambda _k}}}{{{A_k}\ln 2}}\left[ {\int_0^\infty  {\frac{{{{\bar F}_{{\gamma _{{k_{n*}}}}}}\left( z \right)}}{{1 + z}}} dz + \int_0^{\frac{{{a_{m,k}}}}{{{a_{n,k}}}}} {\frac{{\bar F_{{\gamma _{{k_m}}}}}\left( z \right)}{{1 + z}}} dz} \right],
\end{align}
where ${\bar F_{{\gamma _{{k_m}}}}}\left( z \right)$ and ${{{\bar F}_{{\gamma _{{k_{n*}}}}}}\left( z \right)}$ are given by
\begin{align}\label{CDF km}
{\bar F_{{\gamma _{{k_m}}}}}\left( z \right) = &\int _{{r_k}}^\infty  x \exp \left[ { - \frac{{{\sigma ^2}z{x^{{\alpha _k}}}}}{{{P_k}\eta \left( {{a_{m,k}} - {a_{n,k}}z} \right)}}} \right. \nonumber\\
&\left. { - \Theta \left( {\frac{{z{x^{{\alpha _k}}}}}{{{P_k}\eta \left( {{a_{m,k}} - {a_{n,k}}z} \right)}}} \right) + \Lambda \left( x \right)} \right]dx,
\end{align}
and
\begin{align}\label{CDF rn}
{{\bar F}_{{\gamma _{{k_{n*}}}}}}\left( z \right) = \int_{{r_k}}^\infty  {x\exp \left[ {\Lambda \left( x \right) - \frac{{{\sigma ^2}z{r_k}^{{\alpha _k}}}}{{{P_k}\eta {a_{n,k}}}} - \Theta \left( {\frac{{z{r_k}^{{\alpha _k}}}}{{{P_k}\eta {a_{n,k}}}}} \right)} \right]} dx.
\end{align}
\begin{proof}
The proof procedure is similar to the approach of obtaining \eqref{troughput small cell near}, which is detailed introduced in Appendix~C.
\end{proof}
\end{lemma}

\begin{theorem}\label{throughput small cell theorem}
Conditioned on the HPPPs, the achievable ergodic rate of the small cells can be expressed as follows:
\begin{align}\label{throughput small cell}
{\tau _k} = \tau _k^n + \tau _k^f,
\end{align}
where $\tau _k^n$ and $\tau _k^f$ are obtained from \eqref{troughput small cell near} and \eqref{troughput small cell far}.
%\begin{proof}
%Combining the derived results in terms of achievable ergodic rate for both the near user case and the far user case in \eqref{troughput small cell near} and \eqref{troughput small cell far}, we obtain the desired results.
%\end{proof}
\end{theorem}

Note that the derived results in \eqref{throughput small cell} is a double integral form, since even for some special cases, it is challenging to obtain closed form solutions. However, the derived expression is still much more efficient and also more accurate compared to using Monte Carlo simulations, which highly depends on the repeated iterations of random sampling.

\subsection{Ergodic Rate of Macro Cells}
In massive MIMO aided macro cells, the achievable ergodic rate can be significantly improved  due to multiple-antenna array gains, but with more power consumption and high complexity. However, the exact analytical results require high order derivatives of the Laplace transform with the aid of Faa Di Bruno's formula \cite{AndrewsTWC2013}. When the number of antennas goes large, it becomes mathematically intractable to calculate the derivatives due to the unacceptable complexity.   In order to evaluate spectrum efficiency for the whole system,  we provide a tractable lower bound of throughput for macro cells in the following theorem.

\begin{theorem}\label{throughput lower bound theorem}
The lower bound of the achievable ergodic rate of the macro cells can be expressed as follows:
\begin{align}\label{throughput lower bound}
{\tau _{1,L}} = {\log _2}\left( {1 + \frac{{{P_1}{G_M}\eta }}{{N\int_0^\infty  {\left( {{Q_1}\left( x \right) + {\sigma ^2}} \right)} {x^{{\alpha _1}}}{f_{{d_{o,1}}}}\left( x \right)dx}}} \right),
\end{align}
where ${f_{{d_{o,1}}}}\left( x \right)$ is given in \eqref{PDF distance_macro}, ${Q_1}\left( x \right) = \frac{{2{P_1}\eta \pi {\lambda _1}}}{{{\alpha _1} - 2}}{x^{2 - {\alpha _1}}} + \sum\nolimits_{i = 2}^K {2\pi {\lambda _i}\left( {\frac{{{P_i}\eta }}{{{\alpha _i} - 2}}} \right){{\left[ {{\omega _{i,1}}\left( x \right)} \right]}^{2 - {\alpha _i}}}} $, and ${\omega _{i,1}}\left( x \right) = {\left( {\frac{{{a_{n,i}}{{\tilde P}_{i1}}{B_i}N}}{{{G_M}}}} \right)^{\frac{{{\delta _i}}}{2}}}{x^{\frac{1}{{{{\tilde \alpha }_{i1}}}}}}$ is denoted as the nearest distance allowed between the $i$-th tier small cell BS and the typical user that is associated with the macro cell.

\begin{proof}
See Appendix C.
\end{proof}
\end{theorem}

\begin{corollary}\label{proposition:SE lower bound}
If ${\alpha _1} = {\alpha _k} = \alpha $ holds, the lower bound of the achievable ergordic rate of the macro cell is given by in closed-form as
\begin{align}\label{proposition Ergodic Rate macro tight bound exact}
{{\tilde \tau }_{1,L}} = {\log _2}\left( {1 + \frac{{{P_1}{G_M}\eta /N}}{{\psi {{\left( {\pi {b_1}} \right)}^{ - 1}} + {\sigma ^2}\Gamma \left( {\frac{\alpha }{2} + 1} \right){{\left( {\pi {b_1}} \right)}^{ - \frac{\alpha }{2}}}}}} \right),
\end{align}
where $\psi  = \frac{{2{P_1}\eta \pi {\lambda _1}}}{{\alpha  - 2}} + \sum\limits_{i = 2}^K {\left( {\frac{{2\pi {\lambda _i}{P_i}\eta }}{{\alpha  - 2}}} \right)} {\left( {\frac{{{a_{n,i}}{{\tilde P}_{i1}}{B_i}N}}{{{G_M}}}} \right)^{\delta  - 1}}$ and ${b_1} = \sum\limits_{i = 2}^K {{\lambda _i}{{\left( {\frac{{{a_{n,i}}{{\tilde P}_{i1}}{B_i}N}}{{{G_M}}}} \right)}^\delta }}  + {\lambda _1}$.
\begin{proof}
When ${\alpha _1} = {\alpha _k} = \alpha $, \eqref{UA probability macro} can be rewritten as
\begin{align}\label{A1_special}
{{\tilde A}_1} = \frac{{{\lambda _1}}}{{{b_1} }},
\end{align}
Then we have
\begin{align}\label{f macro_special}
{{\tilde f}_{{d_{o,1}}}}\left( x \right) = 2\pi {b_1}x\exp \left( { - \pi {b_1}{x^2}} \right).
\end{align}
By substituting the \eqref{f macro_special} into \eqref{throughput lower bound},  we can obtain
\begin{align}\label{f macro_special 1}
{{\tilde \tau }_{1,L}} = {\log _2}\left( {1 + \frac{{{P_1}{G_M}\eta /N}}{{\int_0^\infty  {\left( {{{\tilde Q}_1}\left( x \right) + {\sigma ^2}} \right)} {x^\alpha }{{\tilde f}_{{d_{o,1}}}}\left( x \right)dx}}} \right),
\end{align}
where ${{\tilde Q}_1}\left( x \right) = \frac{{2{P_1}\eta \pi {\lambda _1}}}{{\alpha  - 2}}{x^{2 - \alpha }} + $ \\$\sum\limits_{i = 2}^K {\left( {\frac{{2\pi {\lambda _i}{P_i}\eta }}{{\alpha  - 2}}} \right)} {\left( {\frac{{{a_{n,i}}{{\tilde P}_{i1}}{B_i}N}}{{{G_M}}}} \right)^{\delta  - 1}}{x^{2 - \alpha }} + {\sigma ^2}$. Then with the aid of \cite[ Eq. (3.326.2)]{gradshteyn}, we obtain the desired closed-form expression as  \eqref{proposition Ergodic Rate macro tight bound exact}. The proof is complete.
\end{proof}
\end{corollary}

\begin{remark}\label{proposition:SE lower bound remark}
The derived results in \eqref{proposition Ergodic Rate macro tight bound exact} demonstrate that the achievable ergordic rate of the macro cell can be enhanced by increasing the
number of antennas at the macro cell BSs. This is because the users in the macro cells can experience larger array gains.
\end{remark}

\subsection{Spectrum Efficiency of the Proposed Hybrid Hetnets}
Based on the analysis of last two subsections, a tractable lower bound of spectrum efficiency can be given  in the following Proposition.
\begin{proposition}\label{SE Hetnets total}
The spectrum efficiency of the proposed hybrid Hetnets is
\begin{align}\label{SE Hetnets}
{\tau _{\mathrm{SE,L}}} ={A_1} N{\tau _{1,L}} + \sum\nolimits_{k = 2}^K {{A_k}{\tau _k}},
\end{align}
where $N{\tau _1}$ and ${{\tau _k}}$ are the lower bound spectrum efficiency of macro cells and the exact spectrum efficiency of the $k$-th tier small cells. Here, ${A_k}$ and ${A_1}$ are obtained from \eqref{UA probability k}  and \eqref{UA probability macro}, ${\tau _k}$ and  ${\tau _{1,L}}$  are obtained from  \eqref{throughput small cell} and \eqref{throughput lower bound}, respectively.
\end{proposition}

\section{Energy Efficiency}
In this section, we proceed to investigate the performance of the proposed hybrid HetNets framework from the perspective of energy efficiency, due to the fact that energy efficiency is an important performance metric for 5G systems.
\subsection{Power Consumption Model}
To calculate the energy efficiency, we first need to model the power consumption parameter of both small cell BSs and macro cell BSs. The power consumption of small cell BSs is given by
\begin{align}\label{power consumption small cell BS}
{P_{i,total}} = {P_{i,static}} + \frac{{{P_i}}}{{{\varepsilon _i}}},
\end{align}
where ${P_{i,static}} $ is the static hardware power consumption of small cell BSs in the $i$-th tier, and $\varepsilon _i$ is the efficiency factor for the power amplifier of small cell BSs in the $i$-th tier.

The power consumption of macro cell BSs is given by
\begin{align}\label{power consumption macro cell BS}
{P_{1,total}} = {P_{1,static}} + \sum\limits_{a = 1}^3 {\left( {{N^a}{\Delta _{a,0}} + {N^{a - 1}}M{\Delta _{a,1}}} \right)}  + \frac{{{P_1}}}{{{\varepsilon _1}}},
\end{align}
where  ${P_{1,static}} $ is the static hardware power consumption of macro cell BSs, $\varepsilon _1$ is the efficiency factor for the power amplifier of macro cell BSs, and $\Delta _{a,0}$ and $\Delta _{a,1}$ are the practical parameters which are depended on the chains of transceivers, precoding, coding/decoding, etc\footnote{The power consumption parameters applied in this treatise are based on an established massive MIMO model proposed in \cite{power2014massive,TWC2015massive}.}.
\subsection{Energy Efficiency of NOMA enhanced Small Cells and Macro Cells}
The energy efficiency is defined as
\begin{align}\label{EE definition}
{\Theta _{\mathrm{EE}}} = \frac{{\mathrm{Total\;data\;rate}}}{{\mathrm{Total\;energy\;consumption}}}.
\end{align}
Therefore, based on \eqref{EE definition} and the power consumption model for small cells that we have provided in \eqref{power consumption small cell BS}, the energy efficiency of the $k$-th tier of NOMA enhanced small cells is expressed as
\begin{align}\label{EE small}
\Theta _{{\mathrm{EE}}}^k= \frac{{{\tau _k}}}{{{P_{k,total}}}},
\end{align}
where ${{\tau _k}}$ is obtained from \eqref{throughput small cell}.

Based on  \eqref{power consumption macro cell BS}   and \eqref{EE definition},  the energy efficiency of macro cell is expressed as
\begin{align}\label{EE macro}
\Theta _{\mathrm{EE}}^1 = \frac{{N{\tau _{1,L}}}}{{{P_{1,total}}}},
\end{align}
where ${{\tau _{1,L}}}$ is obtained from \eqref{throughput lower bound}.
\subsection{Energy Efficiency of the Proposed Hybrid Hetnets}
According to the derived results of energy efficiency of NOMA enhanced small cells and macro cells, we can express the energy efficiency in the following Proposition.
\begin{proposition}\label{EE Hetnets total}
The energy efficiency of the proposed hybrid Hetnets is as follows:
\begin{align}\label{EE HetNets}
\Theta _{\mathrm{EE}}^{\mathrm{Hetnets}} = {A_1}\Theta _{\mathrm{EE}}^1 + \sum\nolimits_{k = 2}^K {{A_k}\Theta _{\mathrm{EE}}^k} ,
\end{align}
where ${A_k}$ and ${A_1}$ are obtained from \eqref{UA probability k}  and \eqref{UA probability macro}, $\Theta _{{\rm{EE}}}^k$  and $\Theta _{\mathrm{EE}}^1$  are obtained from \eqref{EE small} and \eqref{EE macro}.
\end{proposition}

\section{Numerical Results}
\begin{table}[!t]
\centering
\caption{Table of Parameters}
\label{parameter}
\begin{tabular}{|l|l|}
\hline
Monte Carlo simulations repeated  &  ${10^5}$ times \\ \hline
The radius of the plane  &  $10^4$ m \\ \hline
Carrier frequency & $1$~GHz  \\ \hline
The BS density of macro cells & ${\lambda _1} = {\left( {{{500}^2} \times \pi } \right)^{ - 1}}$  \\ \hline
Pass loss exponent & ${\alpha _1} = 3.5$, $\alpha_k=4$  \\ \hline
The noise figure &  $N_f=10$ dB \\ \hline
The noise power & $\sigma^2=-90$ dBm\\ \hline
Static hardware power consumption  &  ${P_{1,total}} = 4$ W, ${P_{i,total}} = 2$ W   \\ \hline
Power amplifier efficiency factor &  ${{\varepsilon _1}}={{\varepsilon _i}}=0.4$ \\ \hline
Precoding power consumption & ${\Delta _{1,0}} = 4.8, {\Delta _{2,0}} = 0$\\ \hline
------  & ${\Delta _{3,0}} = 2.08 \times {10^{ - 8}} $\\ \hline
------ & ${\Delta _{1,1}} = 1, {\Delta _{2,1}} = 9.5 \times {10^{ - 8}}$ \\ \hline
------ & ${\Delta _{3,1}} = 6.25 \times {10^{ - 8}}$ \\ \hline
\end{tabular}
\end{table}
In this section, numerical results are presented to facilitate the performance evaluations of NOMA enhanced hybrid $K$-tier HetNets.  The noise power is $\sigma^2=-170+10 \times {\log _{10}}\left( {BW} \right) + {N_f}$. The power sharing coefficients of NOMA for each tier are same as $a_{m,k}=a_m$ and $a_{n,k}=a_n$ for simplicity. BPCU is short for bit per channel use.  Monte Carlo simulations marked as `$\circ$' are provided to verify the accuracy of our analysis. Table~\ref{parameter} summarizes the the simulation parameters used in this section.

\subsection{User Association Probability and Coverage Probability}

\begin{figure}[t!]
    \begin{center}
        \includegraphics[width=3.4in]{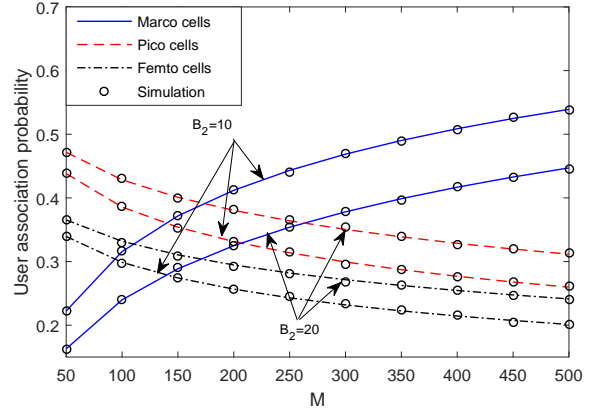}
        \caption{User association probability versus antenna number with different bias factor, with $K=3$, $N=15$, ${P_1} = 40$ dBm, ${P_2} = 30$ dBm and ${P_3} = 20$ dBm, $r_k= 50$ m, $a_m=0.6$, $a_n=0.4$, ${\lambda _2}={\lambda _3}= 20 \times {\lambda _1}$, and $B_{3}=20\times B_{2}$.}
        \label{user asso}
    \end{center}
\end{figure}

\begin{figure}[t!]
    \begin{center}
        \includegraphics[width=3.4in]{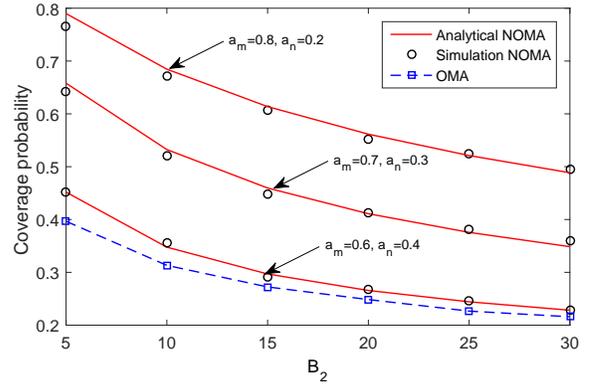}
        \caption{Coverage probability comparison of NOMA and OMA based small cells. $K=2$, $M=200$, $N=15$, ${\lambda _2} = 20 \times {\lambda _1}$, $R_t=R_c=1$ BPCU, $r_k= 10$ m ${P_1} = 40$ dBm, and ${P_2} = 20$ dBm.}
        \label{successful probability of small cells}
    \end{center}
\end{figure}

Fig. \ref{user asso} shows the effect of the number of antennas equipped at each macro BS, $M$, and the bias factor on the user association probability, where the tiers of HetNets are set to be $K=3$, including macro cells and two tiers of small cells. The analytical curves representing small cells  and macro cells are from \eqref{UA probability k} and \eqref{UA probability macro}, respectively.  One can observe that as the number of antennas at each macro BS increases, more users are likely to associate to macro cells. This is because the massive MIMO aided macro cells are capable of providing larger array gain, which in turn enhances the average received power for the connected users. This observation is consistent with \textbf{Remark \ref{User association remark}} in Section III. Another observation is that increasing the bias factor can encourage more users to connect to the small cells, which is an efficient way to extend the coverage of small cells or control the load balance among each tier of HetNets.
Fig. \ref{successful probability of small cells} plots the coverage probability of a typical user associated to the $k$-tier NOMA enhanced small cells versus the bias factor. The solid curves representing the analytical results of NOMA are from  \eqref{successful_probability_typical_final}. One can observe that the coverage probability decreases as the bias factor increases, which means that the unbiased user association  outperforms the biased one, i.e.,  when $B_2=1$, the scenario becomes unbiased user association. This is because by invoking biased user association, users cannot be always associated to the BS which provides the highest received power. But the biased user association is capable of offering more flexibility for users as well as the whole network, especially for the case that cells are fully over load. We also demonstrate that NOMA has superior behavior over OMA scheme\footnote{The OMA benchmark adopted in this treatise is that by dividing the two users in equal time/frequency slots.}.  Actually, the OMA based HetNets scheme has been analytically investigated in the previous research contributions such as \cite{jo2012heterogeneous}, the OMA benchmark adopted in this treatise is generated by numerical approach. It is worth pointing out that power sharing between two NOMA users has a significant effect on coverage probability, and optimizing the power sharing coefficients  can further enlarge the performance gap over OMA based schemes \cite{Yuanwei2016fairness}, which is out of the scope of this paper.

\begin{figure}[t!]
    \begin{center}
        \includegraphics[width=3.4in]{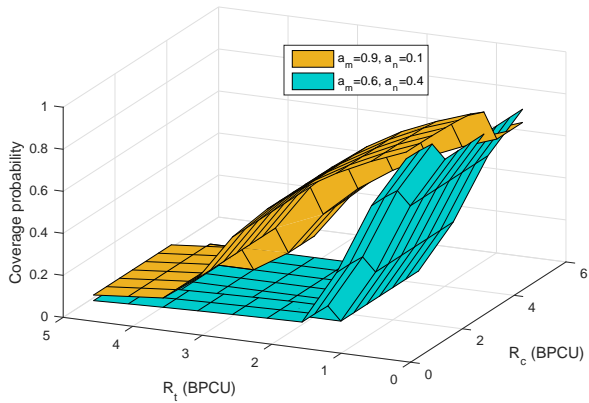}
        \caption{Successful probability of typical user versus targeted rates of $R_t$ and  $R_c$, with $K=2$, $M=200$, $N=15$, ${\lambda _2} = 20 \times {\lambda _1}$, $r_k= 15$ m, $B_2=5$, ${P_1} = 40$ dBm, and ${P_2} = 20$ dBm.}
        \label{successful probability 3D}
    \end{center}
\end{figure}

Fig. \ref{successful probability 3D} plots the coverage probability of a typical user associated to the $k$-tier NOMA enhanced small cells versus both $R_t$ and $R_c$. We observe that there is a cross between these two plotted surfaces, which means that there exists an optimal power sharing allocation scheme for the given targeted rate. In contrast, for fixed power sharing coefficients, e.g., $a_m=0.9, a_n=0.1$, there also exists optimal targeted rates of two users for coverage probability. This figure also illustrates that for inappropriate power and targeted rate selection, the coverage probability is always zero, which also verifies our obtained insights in  \textbf{Remark \ref{coverage small remark}}.

\subsection{Spectrum Efficiency}

\begin{figure}[t!]
    \begin{center}
        \includegraphics[width=3.4in]{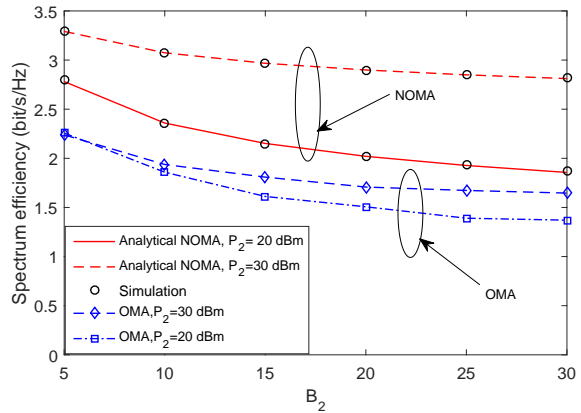}
        \caption{Spectrum efficiency comparison of NOMA and OMA based small cells. $K=2$, $M=200$, $N=15$, $r_k= 50$ m, $a_m=0.6$, $a_n=0.4$, ${\lambda _2} = 20 \times {\lambda _1}$, and ${P_1} = 40$ dBm.}
        \label{ergodic rate of small cells}
    \end{center}
\end{figure}

Fig. \ref{ergodic rate of small cells} plots the spectrum efficiency of small cells with NOMA and OMA versus bias factor, $B_2$, with different transmit powers of small cell BSs, $P_2$.  The curves representing the performance of NOMA enhanced small cells are from \eqref{throughput small cell}. The performance of conventional OMA based small cells is illustrated as a benchmark to demonstrate the effectiveness of our proposed framework.  We observe that the spectrum efficiency of small cells decreases as the bias factor increases. This behavior can be explained as follows: larger bias factor associates more macro users with low SINR to small cells, which in turn  degrades the spectrum efficiency of small cells. It is also worth noting that the performance of NOMA enhanced small cells outperforms the conventional OMA based small cells, which in turn enhances the spectrum efficiency of the whole HetNets.

\begin{figure}[t!]
    \begin{center}
        \includegraphics[width=3.4in]{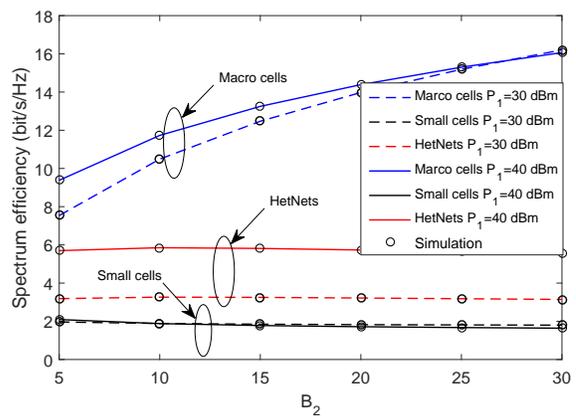}
        \caption{Spectrum efficiency of the proposed framework. $r_k= 50$ m, $a_m=0.6$, $a_n=0.4$, $K=2$, $M=50$, $N=5$, ${P_2} = 20$ dBm, and ${\lambda _2} = 100 \times {\lambda _1}$.}
        \label{SE HetNets}
    \end{center}
\end{figure}
Fig. \ref{SE HetNets} plots the spectrum efficiency of the proposed hybrid HetNets versus bias factor, $B_2$, with different transmit powers, $P_1$. The curves representing the spectrum efficiency of small cells, macro cells and HetNets are from \eqref{SE Hetnets}. We can observe that macro cells can achieve higher spectrum efficiency compared to small cells. This is attributed to the fact that macro BSs are able to serve multiple users simultaneously offering promising array gains to each user, which has been analytically demonstrated in \textbf{Remark \ref{proposition:SE lower bound remark}}. It is also shown that the spectrum efficiency of macro cells improves as the bias factor increases. The reason is again that when more low SINR macro cell users  are associated to small cells, the spectrum efficiency of macro cells can be enhanced.

\subsection{Energy Efficiency}
\begin{figure}[t!]
    \begin{center}
        \includegraphics[width=3.4in]{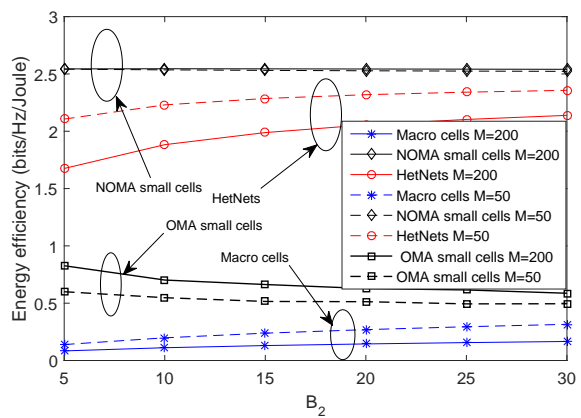}
        \caption{Energy efficiency of the proposed framework. $K=2$, $r_k= 10$ m, $a_m=0.6$, $a_n=0.4$,  $N=15$,  ${P_1} = 30$ dBm, ${P_2} = 20$ dBm, and ${\lambda _2} = 20 \times {\lambda _1}$.}
        \label{EE_HetNets}
    \end{center}
\end{figure}
Fig. \ref{EE_HetNets} plots the energy efficiency of the proposed hybrid HetNets versus the bias factor, $B_2$, with different numbers of transmit antenna of macro cell BSs, $M$. Several observations are as follows: 1) The energy efficiency of the macro cells decrease as the number of antenna increases. Although enlarging the number of antenna at the macro BSs offers a larger array gains, which in turn enhances the spectrum efficiency. Such operations also bring significant power consumption from the baseband signal processing of massive MIMO, which results in decreased energy efficiency. 2) Another observation is that NOMA enhanced small cells can achieve higher energy efficiency than the massive MIMO aided macro cells. It means that from the perspective of energy consumption, densely deploying  BSs in NOMA enhanced small cell is a more effective approach. 3) It is also worth noting that the number of antennas at the macro cell BSs almost has no effect on the energy efficiency of the NOMA enhanced small cells. 4) It also demonstrates that NOMA enhanced small cells has superior performance than conventional OMA based small cells in terms of energy efficiency. Such observations above demonstrate the benefits of the proposed NOMA enhanced hybrid HetNets and  provide insightful guidelines for designing the practical large scale networks.

%\vspace{-0.5cm}

\section{Conclusions}

In this paper, a novel hybrid  HetNets framework has been designed. A flexible  NONA and massive MIMO based user association policy was considered. Stochastic geometry was employed to model the networks and evaluate its performance. Analytical expressions for the coverage probability of NOMA enhanced small cells were derived. It was analytically demonstrated that the inappropriate power allocation among two users will result `always ZERO' coverage probability. Moreover, analytical results for the spectrum efficiency and energy efficiency of the whole network was obtained.  It was interesting to observe that the number of antenna at the macro BSs has weak effects on the energy efficiency of NOMA enhanced small cells.  It has been demonstrated that NOMA enhanced small cells were able to coexist well with the current HetNets structure and were capable of achieving superior performance compared to OMA based small cells. Note that applying NOMA scheme also brings  hardware complexity and processing delay to the existing HetNets structure, which should be taken into considerations.  A promising future direction is to optimize power sharing coefficients among NOMA users to further enhance the performance of the proposed framework.

\section*{Appendix~A: Proof of Lemma \ref{Typical Laplace conventional NOMA small}} \label{Appendix:A}
\renewcommand{\theequation}{A.\arabic{equation}}
\setcounter{equation}{0}

Based on \eqref{SINR small cell kn m near case}, the Laplace transform of the interference from small cell BSs can be expressed as follows:
\begin{align}\label{Laplace_near_typical_1}
&{{\cal L}_{I_{S,k}}}\left( s \right) = {\mathbb{E}_{I_{S,k}}}\left[ {{e^{ - s{I_{S,k}}}}} \right]\nonumber\\
&\mathop  = \limits^{\left( a \right)} {\mathbb{E}_{{\Phi _i}}}\left[ {\sum\nolimits_{i = 2}^K {\prod\limits_{j \in {\Phi _i}\backslash {B_{o,k}}} {{\mathbb{E}_{{g_{j,i}}}}\left[ {{e^{ - s{P_i}{g_{j,i}}\eta d_{j,i}^{ - {\alpha _i}}}}} \right]} } } \right]\nonumber\\
&\mathop  = \limits^{(b)} \exp \left( { - \sum\limits_{i = 2}^K {{\lambda _i}2\pi \int_{{\omega _{i,k}}\left( {{x_0}} \right)}^\infty  {\left( {1 - {\mathbb{E}_{{g_{j,i}}}}\left[ {{e^{ - \frac{{{g_{j,i}}s{P_i}\eta }}{{{r^{{\alpha _i}}}}}}}} \right]} \right)rdr} } } \right)\nonumber\\
&= \exp \left( { - \sum\limits_{i = 2}^K {{\lambda _i}2\pi \int_{{\omega _{i,k}}\left( {{x_0}} \right)}^\infty  {\left( {1 - {{\cal L}_{{g_{j,i}}}}\left( {s{P_i}\eta {r^{ - {\alpha _i}}}} \right)} \right)rdr} } } \right)\nonumber\\
&\mathop  = \limits^{\left( c \right)} \exp \left( { - \sum\limits_{i = 2}^K {{\lambda _i}2\pi \int_{{\omega _{i,k}}\left( {{x_0}} \right)}^\infty  {\left( {1 - {{\left( {1 + s{P_i}\eta {r^{ - {\alpha _i}}}} \right)}^{ - 1}}} \right)rdr} } } \right),
\end{align}
where (a) is resulted from applying Campbell's theorem, (b) is obtained by using the generating-function of PPP, and (c) is obtained by ${{g_{j ,i}}}$ follows exponential distribution with unit mean. By applying \cite[ Eq. (3.194.2)]{gradshteyn}, we can obtain the Laplace transform in an more elegant form in \eqref{Laplace_near_typical_2}. The proof is complete.

\section*{Appendix~B: Proof of Lemma \ref{Typical Laplace conventional NOMA macro}} \label{Appendix:B}
\renewcommand{\theequation}{B.\arabic{equation}}
\setcounter{equation}{0}
Based on \eqref{SINR small cell kn m near case}, the Laplace transform of the interference from macro cell BSs can be expressed as follows:
\begin{align}\label{Lapalace M k 1}
&{\mathcal{L}_{{I_{M,k}}}}\left( s \right) = {\mathbb{E}_{{I_{M,k}}}}\left[ {\exp \left( { - s{\sum _{\ell  \in {\Phi _1}}}\frac{{{P_1}}}{N}{g_{\ell ,1}}L\left( {{d_{\ell ,1}}} \right)} \right)} \right] \nonumber\\
&= {\mathbb{E}_{{\Phi _1}}}\left[ {\prod\limits_{\ell  \in {\Phi _1}} {{\mathbb{E}_{{g_{\ell ,1}}}}} \left[ {\exp \left( { - s\frac{{{P_1}}}{N}{g_{\ell ,1}}\eta d_{\ell ,1}^{ - {\alpha _1}}} \right)} \right]} \right]\nonumber\\
&\mathop  = \limits^{\left( a \right)} \exp \left( { - {\lambda _1}2\pi \int _{{\omega _{i,1}}(x)}^\infty \left( {1 - {\mathbb{E}_{{g_{\ell ,1}}}}\left[ {{e^{ - \frac{{s{P_1}{g_{\ell ,1}}\eta }}{{N{r^{{\alpha _1}}}}}}}} \right]} \right)rdr} \right),
\end{align}
where (a) is obtained with the aid of invoking generating-function of PPP. Recall that the ${g_{\ell ,1}}$ follows Gamma distribution with parameter $\left( {N,1} \right)$. With the aid of Laplace transform for the Gamma distribution, we obtain ${E_{{g_{\ell ,1}}}}\left[ {\exp \left( { - s\frac{{{P_1}}}{N}{g_{\ell ,1}}\eta {r^{ - {\alpha _1}}}} \right)} \right] = {\mathcal{L}_{{g_{\ell ,1}}}}\left( {s\frac{{{P_1}}}{N}\eta {r^{ - {\alpha _1}}}} \right) = {\left( {1 + s\frac{{{P_1}}}{N}\eta {r^{ - {\alpha _1}}}} \right)^{ - N}}$. As such, we can rewrite \eqref{Lapalace M k 1} as
\begin{align}\label{Lapalace M k 2}
&{\mathcal{L}_{{I_{M,k}}}}\left( s \right) = \nonumber\\
&\exp \left( { - {\lambda _1}2\pi \int _{{\omega _{1,k}}\left( x \right)}^\infty \left( {1 - {{\left( {1 + \frac{{s{P_1}\eta }}{{N{r^{{\alpha _1}}}}}} \right)}^{ - N}}} \right)rdr} \right)\nonumber\\
&\mathop  = \limits^{\left( a \right)} \exp \left( { - 2\pi {\lambda _1}\sum\limits_{p = 1}^n {
n\choose
p} {{\left( {\frac{{s\eta {P_1}}}{N}} \right)}^p}\int_{{\omega _{1,k}}\left( {{x_0}} \right)}^\infty  {\frac{{{r^{ - {\alpha _1}p + 1}}}}{{{{\left( {1 + \frac{{s\eta {P_1}}}{{{r^\alpha }N}}} \right)}^N}}}} dr} \right)\nonumber\\
&\mathop  = \limits^{\left( b \right)} \exp \left[ { - \pi {\lambda _1}{\delta _1}{{\left( {\frac{{s\eta {P_1}}}{N}} \right)}^{{\delta _1}}}\sum\limits_{p = 1}^N {
N\choose
p} {{\left( { - 1} \right)}^{{\delta _1} - p}}} \right.\nonumber\\
&\left. {\,\,\,\,\,\,\,\,\,\, \times \int_0^{ - {\omega _{1,k}}{{\left( x \right)}^{ - \alpha }}s\eta {P_1}/N} {\frac{{{t^{p - {\delta _1} - 1}}}}{{{{\left( {1 - t} \right)}^N}}}} dt} \right],
\end{align}
where $(a)$ is obtained by applying binomial expression and after some mathematical manipulations, and $(b)$ is obtained by using $t =  - s\eta {r^{ - {\alpha _1}}}{P_1}/N$. Based on \cite[ Eq. (8.391)]{gradshteyn}, we can obtain the Laplace transform of ${I_{M,k}}$ as given in \eqref{Lapalace M k final}. The proof is complete.

\section*{Appendix~C: Proof of Lemma \ref{theorem:small near}} \label{Appendix:C}
\renewcommand{\theequation}{C.\arabic{equation}}
\setcounter{equation}{0}
For the near user case in small cells, the achievable ergodic rate in the $k$-th tier can be expressed as

\begin{align}\label{Ergodic Rate small cell Expetation}
\tau _k^n &= E\left\{ {{{\log }_2}\left( {1 + {\gamma _{{k_{m^*}}}}} \right) + {{\log }_2}\left( {1 + {\gamma _{{k_n}}}} \right)} \right\}\nonumber\\
&= \frac{1}{{\ln 2}}\int_0^\infty  {\frac{{{{\bar F}_{{\gamma _{{k_{m*}}}}}}\left( z \right)}}{{1 + z}}} dz + \frac{1}{{\ln 2}}\int_0^\infty  {\frac{{{{\bar F}_{{\gamma _{{k_n}}}}}\left( z \right)}}{{1 + z}}} dz.
\end{align}
We need to obtain the expressions for ${{\bar F}_{{\gamma _{{k_n}}}}}\left( z \right)$ first.  Based on \eqref{SINR small cell kn near case}, we can obtain
\begin{align}\label{CDF km 1}
&{{\bar F}_{{\gamma _{{k_n}}}}}\left( z \right)= \int_0^{{r_k}} {\Pr \left[ {\frac{{{a_{n,k}}{P_k}{g_{o,k}}\eta {x^{ - {\alpha _k}}}}}{{{I_{M,k}} + {I_{S,k}} + {\sigma ^2}}} > z} \right]} {f_{{d_{o,k}}}}\left( x \right)dx\nonumber\\
= & \int_0^{{r_k}} {\exp \left( { - \frac{{{\sigma ^2}z{x^{{\alpha _k}}}}}{{{a_{n,k}}{P_k}\eta }}} \right)} {\mathcal{L}_{{I_{k}}}}\left( {\frac{{z{x^{{\alpha _k}}}}}{{{a_{n,k}}{P_k}\eta }}} \right){f_{{d_{o,k}}}}\left( x \right)dx,
\end{align}

%Note that for the case $z \ge \frac{{{a_m}}}{{{a_n}}}$, it is easy to observe that ${F_{r,{k_m}}}\left( z \right)=1$. For the case $z \le \frac{{{a_m}}}{{{a_n}}}$, we can obtain
%\begin{align}\label{CDF km 2}
%&{F_{r,{k_m}}}\left( z \right) = \int_0^\infty  {\Pr \left[ {{g_{o,k}} \le \frac{{\left( {{I_{M,k}} + {I_{S,k}} + {\sigma ^2}} \right)z{x^{{\alpha _k}}}}}{{{P_k}\eta \left( {{a_m} - {a_n}z} \right)}}} \right]} {f_{{d_{o,{k_m}}}}}\left( x \right)dx \nonumber\\
%&= 1 - \int_0^{{r_k}} {\exp \left( { - \frac{{{\sigma ^2}z{x^{{\alpha _k}}}}}{{{a_{n,k}}{P_k}\eta }}} \right)} {L_{{I_{k*}}}}\left( {\frac{{z{x^{{\alpha _k}}}}}{{{a_{n,k}}{P_k}\eta }}} \right){f_{{d_{o,k}}}}\left( x \right)dx,
%\end{align}

By combining \eqref{Lapalace M k final} and \eqref{Laplace_near_typical_2}, we can obtain the Laplace transform of ${{I_{k*}}}$ as ${{\cal L}_{{I_{k*}}}}\left( s \right) = \exp \left( { - \Theta \left( s \right)} \right)$, where ${\Theta \left( {s} \right)}$ is given in \eqref{laplace inter}. By plugging \eqref{PDF distance_k} and ${\mathcal{L}_{{I_{k*}}}}\left( s \right)$ into \eqref{CDF km 1}, we obtain the complete cumulative distribution function (CCDF) of ${{\gamma _{{k_n}}}}$ in \eqref{CDF kn}.
In the following, we turn to our attention to derive the  CCDF of ${{\gamma _{{k_{m*}}}}}$. Based on \eqref{SINR small cell rm near case}, we can obtain ${{{\bar F}_{{\gamma _{{k_{m*}}}}}}\left( z \right)}$ as
%\vspace{-0.2cm}
\begin{align}\label{CDF rm 1}
&{{{\bar F}_{{\gamma _{{k_{m*}}}}}}\left( z \right)}= \int_0^{{r_k}} {{f_{{d_{o,k}}}}\left( x \right)}  \times\nonumber\\
&\Pr \left[ {\left( {{a_{m,k}} - {a_{n,k}}z} \right){g_{o,k}} > \frac{{\left( {{I_{M,k}} + {I_{S,k}} + {\sigma ^2}} \right)z}}{{{P_k}\eta {r_k}^{ - {\alpha _k}}}}} \right]dx.
\end{align}
Note that for the case $z \ge \frac{{{a_{m,k}}}}{{{a_{n,k}}}}$, it is easy to observe that ${{{\bar F}_{{\gamma _{{k_{m*}}}}}}\left( z \right)}=0$. For the case $z \le \frac{{{a_{m,k}}}}{{{a_{n,k}}}}$, following the similar procedure of deriving \eqref{CDF kn}, we can obtain the ergodic rate of the existing user for the near user case as \eqref{CDF rm}. The proof is complete.

%\begin{align}\label{CDF rm 2}
%{F_{{k_{m^*}}}}\left( z \right) =& 1 - \exp \left( { - \frac{{{\sigma ^2}z{r_k}^{{\alpha _k}}}}{{\left( {{a_{m,k}} - {a_{n,k}}z} \right){P_k}\eta }}} \right)\nonumber\\
%&\times{L_{{I_{k*}}}}\left( {\frac{{z{r_k}^{{\alpha _k}}}}{{\left( {{a_{m,k}} - {a_{n,k}}z} \right){P_k}\eta }}} \right).
%\end{align}
%Plugging ${L_{{I_{k*}}}}\left( s \right)$ into \eqref{CDF rm 2}, combining the two cases aforementioned, we obtain
%\begin{align}\label{CDF rm 2}
%&{F_{{k_{m^*}}}}\left( z \right) = 1 - U\left( {\frac{{{a_{m,k}}}}{{{a_{n,k}}}} - z} \right) \times \nonumber\\
%&\exp \left[ { - \frac{{{\sigma ^2}z{r_k}^{{\alpha _k}}}}{{\left( {{a_{m,k}} - {a_{n,k}}z} \right){P_k}\eta }}-\Theta \left( {\frac{{z{r_k}^{{\alpha _k}}}}{{\left( {{a_{m,k}} - {a_{n,k}}z} \right){P_k}\eta }},{r_k}} \right)} \right],
%\end{align}
%where $U\left( {\cdot} \right)$ is the unit step function. Substituting \eqref{CDF rm 2} and the \eqref{CDF kn} into \eqref{Ergodic Rate small cell Expetation}, we obtain the ergodic rate of small cell for the near user case. The proof is complete.

\section*{Appendix~D: Proof of Theorem \ref{throughput lower bound theorem}} \label{Appendix:D}
\renewcommand{\theequation}{D.\arabic{equation}}
\setcounter{equation}{0}

With the aid of Jensen's inequality, we can obtain the lower bound of the achievable ergodic rate of the macro cells as
\begin{align}\label{Ergodic Rate macro 1}
\mathbb{E}\left\{ {{{\log }_2}\left( {1 + {\gamma _{r,1}}} \right)} \right\} \ge \tau_{1,L}={\log _2}\left( {1 + {{\left( {\mathbb{E}\left\{ {{{\left( {{\gamma _{r,1}}} \right)}^{ - 1}}} \right\}} \right)}^{ - 1}}} \right)
\end{align}
By invoking the law of large numbers, we have ${h_{o,1}} \approx {G_M}$. Then based on \eqref{SINR Macro}, $\tau_{1,L}$ can be approximated  as follows:
\begin{align}\label{Ergodic Rate macro 2}
&\mathbb{E}\left\{ {{{\left( {{\gamma _{r,1}}} \right)}^{ - 1}}} \right\} \approx \frac{N}{{{P_1}{G_M}\eta }}\mathbb{E}\left\{ {\left( {{I_{M,1}} + {I_{S,1}} + {\sigma ^2}} \right){x^{{\alpha _1}}}} \right\}\nonumber\\
&= \frac{N}{{{P_1}{G_M}\eta }}\int_0^\infty  {\left( {\mathbb{E}\left\{ {\left. {{I_{M,1}} + {I_{S,1}}} \right|{d_{o,1}} = x} \right\} + {\sigma ^2}} \right)}\nonumber\\
&\times {x^{{\alpha _1}}}{f_{{d_{o,1}}}}\left( x \right)dx.
\end{align}
We turn to our attention to the expectation, denoting ${Q_1}\left( x \right) = \mathbb{E}\left\{ {\left. {{I_{M,1}} + {I_{S,1}}} \right|{d_{o,1}} = x} \right\}$, with the aid of Campbell's Theorem, we obtain
\begin{align}\label{Ergodic Rate macro Expetation}
&{Q_1}\left( x \right) = \mathbb{E}\left\{ {\left. {{\sum _{\ell  \in {\Phi _1}\backslash {B_{o,1}}}}\frac{{{P_1}}}{N}{h_{\ell ,1}}L\left( {{d_{\ell ,1}}} \right)} \right|{d_{o,1}} = x} \right\}\nonumber\\
& +  \mathbb{E}\left\{ {\left. {\sum\nolimits_{i = 2}^K {\sum\nolimits_{j \in {\Phi _i}} {{P_i}{h_{j,i}}L\left( {{d_{j,i}}} \right)} } } \right|{d_{o,1}} = x} \right\}\nonumber\\
&= \frac{{2{P_1}\eta \pi {\lambda _1}}}{{{\alpha _1} - 2}}{x^{2 - {\alpha _1}}} + \sum\limits_{i = 2}^K {2\pi {\lambda _i}\left( {\frac{{{P_i}\eta }}{{{\alpha _i} - 2}}} \right)} {\left[ {{\omega _{i,1}}\left( x \right)} \right]^{2 - {\alpha _i}}},
\end{align}
We first calculate the first part of \eqref{Ergodic Rate macro Expetation} as
\begin{align}\label{Ergodic Rate macro Expetation_part 1}
&\mathbb{E}\left\{ {\left. {{\sum _{\ell  \in {\Phi _1}\backslash {B_{o,1}}}}\frac{{{P_1}}}{N}{h_{\ell ,1}}L\left( {{d_{\ell ,1}}} \right)} \right|{d_{o,1}} = x} \right\}\nonumber\\
\mathop  = \limits^{\left( a \right)} & \frac{{{P_1}}}{N}\eta \mathbb{E}\left\{ {{h_{\ell ,1}}} \right\}{\lambda _1}\int_R {{r^{ - {\alpha _1}}}} dr\nonumber\\
\mathop  = \limits^{\left( b \right)} &2\pi {P_1}\eta {\lambda _1}\int_x^\infty  {{r^{1 - {\alpha _1}}}} dr\nonumber\\
= &\frac{{2{P_1}\eta \pi {\lambda _1}}}{{{\alpha _1} - 2}}{x^{2 - {\alpha _1}}},
\end{align}
where $(a)$ is obtained by applying Campbell's theorem, and $(b)$ is obtained since the expectation of ${{h_{\ell ,1}}}$ is $N$.
Then we turn to our attention to the second part of \eqref{Ergodic Rate macro Expetation}, with using the similar approach, we obtain
\begin{align}\label{Ergodic Rate macro Expetation_part 2}
&\mathbb{E}\left\{ {\left. {\sum\nolimits_{i = 2}^K {\sum\nolimits_{j \in {\Phi _i}} {{P_i}{h_{j,i}}L\left( {{d_{j,i}}} \right)} } } \right|{d_{o,1}} = x} \right\}\nonumber\\
=& \sum\limits_{i = 2}^K {\left( {\frac{{2\pi {\lambda _i}{P_i}\eta }}{{{\alpha _i} - 2}}} \right)} {\left[ {{\omega _{i,1}}\left( x \right)} \right]^{2 - {\alpha _i}}}.
\end{align}
By substituting \eqref{Ergodic Rate macro Expetation_part 1} and \eqref{Ergodic Rate macro Expetation_part 2} into \eqref{Ergodic Rate macro 2}, we obtain the desired results in \eqref{throughput lower bound}. The proof is complete.

\bibliographystyle{IEEEtran}
\bibliography{mybib}

\begin{thebibliography}{10}
\providecommand{\url}[1]{#1}
\csname url@samestyle\endcsname
\providecommand{\newblock}{\relax}
\providecommand{\bibinfo}[2]{#2}
\providecommand{\BIBentrySTDinterwordspacing}{\spaceskip=0pt\relax}
\providecommand{\BIBentryALTinterwordstretchfactor}{4}
\providecommand{\BIBentryALTinterwordspacing}{\spaceskip=\fontdimen2\font plus
\BIBentryALTinterwordstretchfactor\fontdimen3\font minus
  \fontdimen4\font\relax}
\providecommand{\BIBforeignlanguage}[2]{{%
\expandafter\ifx\csname l@#1\endcsname\relax
\typeout{** WARNING: IEEEtran.bst: No hyphenation pattern has been}%
\typeout{** loaded for the language `#1'. Using the pattern for}%
\typeout{** the default language instead.}%
\else
\language=\csname l@#1\endcsname
\fi
#2}}
\providecommand{\BIBdecl}{\relax}
\BIBdecl

\bibitem{yuanwei2016Hetnets}
Y.~Liu, Z.~Qin, M.~Elkashlan, Y.~Gao, and N.~Arumugam, ``Non-orthogonal
  multiple access in massive {MIMO} aided heterogeneous networks,'' in
  \emph{Proc. of Global Commun. Conf. (GLOBECOM)}, Washington D.C, USA, Dec.
  2016.

\bibitem{CiscoVNI}
Cisco, ``Cisco visual networking index: {G}lobal mobile data traffic forecast
  update 2014-2019 white paper,'' Dec. 2015.

\bibitem{IOT2015}
A.~Al-Fuqaha, M.~Guizani, M.~Mohammadi, M.~Aledhari, and M.~Ayyash, ``Internet
  of things: A survey on enabling technologies, protocols, and applications,''
  \emph{{IEEE} Commun. Surveys Tutorials}, vol.~17, no.~4, pp. 2347--2376,
  Fourth quarter 2015.

\bibitem{Dai2015NOMA}
L.~Dai, B.~Wang, Y.~Yuan, S.~Han, C.~l.~I, and Z.~Wang, ``Non-orthogonal
  multiple access for {5G}: solutions, challenges, opportunities, and future
  research trends,'' \emph{{IEEE} Commun. Mag.}, vol.~53, no.~9, pp. 74--81,
  Sep. 2015.

\bibitem{shirvanimoghaddam2016massive}
M.~Shirvanimoghaddam, M.~Dohler, and S.~Johnson, ``Massive non-orthogonal
  multiple access for cellular {IoT}: Potentials and limitations,'' \emph{arXiv
  preprint arXiv:1612.00552}.

\bibitem{Zhijin2017modulation}
Y.~Cai, Z.~Qin, F.~Cui, G.~Y. Li, and J.~A. McCann, ``Modulation and multiple
  access for {5G} networks,'' \emph{arXiv preprint arXiv:1702.07673}, 2017.

\bibitem{Zhiguo2015Mag}
Z.~Ding, Y.~Liu, J.~Choi, Q.~Sun, M.~Elkashlan, C.-L. I, and H.~V. Poor,
  ``Application of non-orthogonal multiple access in {LTE} and {5G} networks,''
  \emph{{IEEE} Commun. Mag.}, Feb. 2017.

\bibitem{Saito:2013}
Y.~Saito, Y.~Kishiyama, A.~Benjebbour, T.~Nakamura, A.~Li, and K.~Higuchi,
  ``Non-orthogonal multiple access ({NOMA}) for cellular future radio access,''
  in \emph{Proc. Vehicular Technology Conference (VTC Spring)}, June Dresden,
  Germany, Jun. 2013, pp. 1--5.

\bibitem{ding2014performance}
Z.~Ding, Z.~Yang, P.~Fan, and H.~V. Poor, ``On the performance of
  non-orthogonal multiple access in {5G} systems with randomly deployed
  users,'' \emph{{IEEE} Signal Process. Lett.}, vol.~21, no.~12, pp.
  1501--1505, Dec. 2014.

\bibitem{Timotheou:2015}
S.~Timotheou and I.~Krikidis, ``Fairness for non-orthogonal multiple access in
  5{G} systems,'' \emph{{IEEE} Signal Process. Lett.}, vol.~22, no.~10, pp.
  1647--1651, Oct. 2015.

\bibitem{Jinho:2015}
J.~Choi, ``Minimum power multicast beamforming with superposition coding for
  multiresolution broadcast and application to {NOMA} systems,'' \emph{{IEEE}
  Trans. Commun.}, vol.~63, no.~3, pp. 791--800, Mar. 2015.

\bibitem{andrews2014will}
J.~G. Andrews, S.~Buzzi, W.~Choi, S.~V. Hanly, A.~Lozano, A.~C. Soong, and
  J.~C. Zhang, ``What will {5G} be?'' \emph{{IEEE} J. Sel. Areas Commun.},
  vol.~32, no.~6, pp. 1065--1082, 2014.

\bibitem{Andrews2008Femto}
V.~Chandrasekhar, J.~G. Andrews, and A.~Gatherer, ``Femtocell networks: a
  survey,'' \emph{{IEEE} Commun. Mag.}, vol.~46, no.~9, pp. 59--67, Sep. 2008.

\bibitem{Xie2016access}
H.~Xie, F.~Gao, and S.~Jin, ``An overview of low-rank channel estimation for
  massive {MIMO} systems,'' \emph{IEEE Access}, vol.~4, pp. 7313--7321, 2016.

\bibitem{Adhikary2015Hetnets}
A.~Adhikary, H.~S. Dhillon, and G.~Caire, ``Massive-{MIMO} meets hetnet:
  Interference coordination through spatial blanking,'' \emph{{IEEE} J. Sel.
  Areas Commun.}, vol.~33, no.~6, pp. 1171--1186, Jun. 2015.

\bibitem{ye2015user}
Q.~Ye, O.~Y. Bursalioglu, H.~C. Papadopoulos, C.~Caramanis, and J.~G. Andrews,
  ``User association and interference management in massive {MIMO} hetnets,''
  \emph{{IEEE} Trans. Commun.}, vol.~64, no.~5, pp. 2049--2065, May 2016.

\bibitem{chiu2013stochastic}
S.~N. Chiu, D.~Stoyan, W.~S. Kendall, and J.~Mecke, \emph{Stochastic geometry
  and its applications}.\hskip 1em plus 0.5em minus 0.4em\relax John Wiley \&
  Sons, 2013.

\bibitem{jo2012heterogeneous}
H.-S. Jo, Y.~J. Sang, P.~Xia, and J.~G. Andrews, ``Heterogeneous cellular
  networks with flexible cell association: A comprehensive downlink {SINR}
  analysis,'' \emph{{IEEE} Trans. Wireless Commun.}, vol.~11, no.~10, pp.
  3484--3495, 2012.

\bibitem{Dhillon2014MIMOHetnets}
A.~K. Gupta, H.~S. Dhillon, S.~Vishwanath, and J.~G. Andrews, ``Downlink
  multi-antenna heterogeneous cellular network with load balancing,''
  \emph{{IEEE} Trans. Commun.}, vol.~62, no.~11, pp. 4052--4067, Nov. 2014.

\bibitem{Wen2016Hetnets}
W.~Liu, S.~Jin, C.~K. Wen, M.~Matthaiou, and X.~You, ``A tractable approach to
  uplink spectral efficiency of two-tier massive {MIMO} cellular hetnets,''
  \emph{{IEEE} Commun. Lett.}, vol.~20, no.~2, pp. 348--351, Feb. 2016.

\bibitem{Anqi2015Hetnets}
A.~He, L.~Wang, M.~Elkashlan, Y.~Chen, and K.~K. Wong, ``Spectrum and energy
  efficiency in massive {MIMO} enabled hetnets: A stochastic geometry
  approach,'' \emph{{IEEE} Commun. Lett.}, vol.~19, no.~12, pp. 2294--2297,
  Dec. 2015.

\bibitem{yuanwei_JSAC_2015}
Y.~Liu, Z.~Ding, M.~Elkashlan, and H.~V. Poor, ``Cooperative non-orthogonal
  multiple access with simultaneous wireless information and power transfer,''
  \emph{{IEEE} J. Sel. Areas Commun.}, vol.~34, no.~4, April 2016.

\bibitem{Zhiguo2016General}
Z.~Ding, R.~Schober, and H.~V. Poor, ``A general {MIMO} framework for {NOMA}
  downlink and uplink transmission based on signal alignment,'' \emph{{IEEE}
  Trans. Wireless Commun.}, vol.~15, no.~6, pp. 4438--4454, June 2016.

\bibitem{Yuanwei2017TWC}
Y.~Liu, Z.~Qin, M.~Elkashlan, Y.~Gao, and L.~Hanzo, ``Enhancing the physical
  layer security of non-orthogonal multiple access in large-scale networks,''
  \emph{{IEEE} Trans. Wireless Commun.}, vol.~16, no.~3, pp. 1656--1672, Mar.
  2017.

\bibitem{Zhiguo2016mmwave}
Z.~Ding, P.~Fan, and H.~V. Poor, ``Random beamforming in millimeter-wave {NOMA}
  networks,'' \emph{{IEEE} Access}, to apprear in 2017.

\bibitem{ding2015mimo}
Z.~Ding, F.~Adachi, and H.~V. Poor, ``The application of {MIMO} to
  non-orthogonal multiple access,'' \emph{{IEEE} Trans. Wireless Commun.},
  vol.~15, no.~1, pp. 537--552, Jan. 2015.

\bibitem{Yuanwei2016fairness}
Y.~Liu, M.~Elkashlan, Z.~Ding, and G.~K. Karagiannidis, ``Fairness of user
  clustering in {MIMO} non-orthogonal multiple access systems,'' \emph{{IEEE}
  Commun. Lett.}, vol.~20, no.~7, pp. 1465--1468, July 2016.

\bibitem{power2014massive}
E.~Bjornson, L.~Sanguinetti, J.~Hoydis, and M.~Debbah, ``Designing multi-user
  {MIMO} for energy efficiency: When is massive {MIMO} the answer?'' in
  \emph{Proc. Wireless Commun. and Networking Conf. (WCNC)}, Apr. 2014, pp.
  242--247.

\bibitem{huh2012network}
H.~Huh, A.~M. Tulino, and G.~Caire, ``Network {MIMO} with linear zero-forcing
  beamforming: Large system analysis, impact of channel estimation, and
  reduced-complexity scheduling,'' \emph{{IEEE} Trans. Inf. Theory}, vol.~58,
  no.~5, pp. 2911--2934, 2012.

\bibitem{3GPP}
3rd Generation Partnership Project~({3GPP}), ``Study on downlink multiuser
  superposition transmission for {LTE},'' Mar. 2015.

\bibitem{hosseini2014large}
K.~Hosseini, W.~Yu, and R.~S. Adve, ``Large-scale {MIMO} versus network {MIMO}
  for multicell interference mitigation,'' vol.~8, no.~5, pp. 930--941, 2014.

\bibitem{steele1999mobile}
R.~Steele and L.~Hanzo, \emph{Mobile Radio Communications: Second and Third
  Generation Cellular and {WATM} Systems: 2nd}, 1999.

\bibitem{gradshteyn}
I.~S. Gradshteyn and I.~M. Ryzhik, \emph{Table of Integrals, Series and
  Products}, 6th~ed.\hskip 1em plus 0.5em minus 0.4em\relax New York, NY, USA:
  Academic Press, 2000.

\bibitem{liu2015secure}
Y.~Liu, L.~Wang, S.~Zaidi, M.~Elkashlan, and T.~Duong, ``Secure {D2D}
  communication in large-scale cognitive cellular networks: A wireless power
  transfer model,'' \emph{{IEEE} Trans. Commun.}, vol.~64, no.~1, pp. 329--342,
  Jan. 2016.

\bibitem{AndrewsTWC2013}
H.~S. Dhillon, M.~Kountouris, and J.~G. Andrews, ``Downlink {MIMO} hetnets:
  Modeling, ordering results and performance analysis,'' \emph{{IEEE} Trans.
  Wireless Commun.}, vol.~12, no.~10, pp. 5208--5222, Oct. 2013.

\bibitem{TWC2015massive}
E.~Bjornson, L.~Sanguinetti, J.~Hoydis, and M.~Debbah, ``Optimal design of
  energy-efficient multi-user {MIMO} systems: Is massive {MIMO} the answer?''
  \emph{{IEEE} Trans. Wireless Commun.}, vol.~14, no.~6, pp. 3059--3075, Jun.
  2015.

\end{thebibliography}

\end{document}